\documentclass[sigconf,nonacm]{acmart}
\setcopyright{none}
\usepackage{mathtools}
\ifPDFTeX
  \usepackage{bm}
\else
  \newcommand{\bm}[1]{\symbf{#1}}
\fi
\usepackage{algorithm}
\usepackage{algorithmic}
\usepackage{tabularx}
\usepackage[utf8]{inputenc}
\usepackage{url}
\usepackage{multirow}
\usepackage{colortbl}
\newtheorem*{proposition*}{Proposition}
\newtheorem*{theorem*}{Theorem}
\usepackage{thmtools}
\usepackage[most]{tcolorbox}
\usepackage{placeins}
\usepackage{soul}
\usepackage{amsthm}
\usepackage{enumitem}

\newcommand*{\here}{}

\definecolor{EInk}{HTML}{245FA4}
\definecolor{ResultBlue}{HTML}{486581}
\newtcolorbox{ResultBox}{enhanced,breakable,
  colback=ResultBlue!3!white,colframe=ResultBlue!70!white,coltext=black,
  boxrule=0.5pt,arc=1pt,left=4pt,right=4pt,top=3pt,bottom=3pt,
  boxsep=0pt,before skip=4pt,after skip=4pt,
  before upper={\setlength{\parskip}{3pt}}}

\providecommand{\dd}[1]{\,{\color{gray}\scriptsize #1}}

\newcolumntype{R}{>{\raggedleft\arraybackslash}p{2.9em}}
\newtheorem{theorem}{Theorem}
\newtheorem{definition}{Definition}
\newtheorem{proposition}{Proposition}
\newtheorem{remark}{Remark}
\newtheorem{lemma}{Lemma}
\newcommand{\oomit}[1]{}

\title{Anytime-Valid Runtime Monitoring of Learned Barrier Certificates}

\author{
    Zhipeng Cao\textsuperscript{\rm 1,2},
    Taoran Wu\textsuperscript{\rm 1,3},
    Weiru Cui\textsuperscript{\rm 1,3},
    Luke Ong\textsuperscript{\rm 4},
    Bai Xue\textsuperscript{\rm 1,2}
}
\affiliation{
  \institution{
    \textsuperscript{\rm 1}Institute of Software, Chinese Academy of Sciences \\
    \textsuperscript{\rm 2}School of Advanced Interdisciplinary Sciences,
University of Chinese Academy of Sciences \\
    \textsuperscript{\rm 3}University of Chinese Academy of Sciences \\
    \textsuperscript{\rm 4}College of Computing and Data Science, Nanyang Technological University
  }
  \city{}
  \country{}
}
\renewcommand{\shortauthors}{Zhipeng Cao et al.}

\begin{document}

\hypersetup{
  pdfauthor={Zhipeng Cao, Taoran Wu, Weiru Cui, Luke Ong, Bai Xue}
}

\begin{abstract}
Learned barrier certificates provide a principled approach to establish the safety of stochastic systems, but their validity may deteriorate during deployment when the underlying dynamics differ from those encountered during training. We propose an anytime-valid runtime monitoring framework that audits learned barrier certificates using observed closed-loop trajectories. Specifically, we formulate the conditional barrier-drift condition as an episode-wise sequential hypothesis test and construct a nonnegative test supermartingale from observed barrier increments using a data-driven adaptive betting strategy. This yields finite-sample and false-alarm rate control under arbitrary within-episode stopping times. To exploit a prescribed finite deployment horizon, we further introduce a deterministic time-varying scaling of the monitoring boundary. For scaling factors $a\geq1$, this makes the monitor more conservative at early time steps while retaining the level-$\alpha$ false-alarm guarantee. We also establish a weighted false-alarm budget and stepwise cumulative false-alarm bounds for all $a>0$, characterizing the trade-off between early detection and false-alarm control. The monitor detects statistical evidence against the deployed barrier condition rather than directly predicting unsafe entry. Finally, we evaluate the framework under valid and invalid barrier certificates, compare it with direct barrier monitoring, and investigate monitoring-guided certificate refinement. In an additional closed-loop evaluation on a learned vision-language-action (VLA) policy, coupling the monitors with a predefined runtime intervention reduces unsafe rollouts, with an average per-step overhead of only $0.43$ ms for barrier evaluation and monitoring.
\end{abstract}


\maketitle

\section{Introduction}
Safety is a critical requirement for learning-enabled autonomous systems operating in uncertain and evolving environments. In robotics, autonomous driving, and cyber-physical infrastructure, even rare failures can be catastrophic, making purely data-driven validation insufficient for deployment~\cite{chougule2023comprehensive,uesato2018rigorous,rao2026geometryflowmatchinguncertaintycostfree}.
Consider an autonomous drone navigating a cluttered warehouse whose floor plan changes over time. A barrier trained in simulation may provide a probabilistic safety guarantee under the assumed system dynamics and disturbance model, but unmodeled disturbances or environmental changes may invalidate the deployed safety certificate. 
This motivates runtime monitoring of the deployed certificate.

Barrier certificates~\cite{prajna2004safety,Prajna2007} certify safety through a scalar function $B(x)$ whose expected value does not increase along trajectories, offering a scalable alternative to reachability analysis. In practice, however, the barrier $B$ is learned from data, which introduces a fundamental verification gap: finite datasets provide incomplete coverage, simulation-to-reality mismatch induces distribution shift, and black-box models resist formal verification. In stochastic environments, iterative test-then-train refinement on counterexamples also cannot guarantee global validity of the updated certificate. Rather than pursuing complete verification, we shift the focus to \emph{runtime monitoring} and \emph{alarm-triggered intervention}.
The key idea is to continuously audit the deployed barrier using observed closed-loop trajectories and detect statistical evidence that the certificate is no longer consistent with the underlying dynamics. 
The monitor accumulates statistical evidence against the barrier’s conditional drift condition and converts this evidence into an alarm for runtime intervention. Collected interaction data then support subsequent barrier or policy updates.

To realize this, we formulate runtime barrier auditing as an anytime-valid sequential  hypothesis test based on  e-processes (test supermartingales) and predictable plug-in betting strategies \cite{ramdas2023game,shafer2011test,vovk2021values,betting2021,waudbysmith2022estimatingmeansboundedrandom}. Observed barrier-value increments are converted into betting factors whose product forms a nonnegative test supermartingale. Consequently, the monitor retains finite-sample false-alarm control under arbitrary stopping times within an episode, allowing evidence to be assessed continuously without sacrificing the prescribed statistical validity. A finite deployment horizon also provides useful temporal structure for monitoring. Inspired by fi\-nite-horizon safety verification \cite{xue2025finite}, we introduce a deterministic time-varying scaling of the monitoring boundary to tailor the monitor to a prescribed deployment horizon. The scaling parameter $a$ controls the temporal preference of the monitor: for $a\geq1$, the boundary is more conservative at early time steps and gradually approaches the standard anytime-valid threshold as the horizon is reached, whereas $0<a<1$ favors earlier detection. We further establish a weighted false-alarm budget and derive stepwise cumulative false-alarm bounds for all $a>0$, thereby characterizing the trade-off between early detection and false-alarm control. In particular, the proposed scaling preserves the level-$\alpha$ false-alarm guarantee for $a\geq1$, while providing an explicit bound on the increased false-alarm probability for $0<a<1$.


The contributions of this work are summarized as follows.
\begin{enumerate}
    \item \textbf{Anytime-valid barrier auditing:}  We formulate the conditional drift condition underlying a stochastic barrier certificate as an episode-wise sequential hypothesis test and construct an e-process from observed barrier increments. The  monitor provides finite-sample level-$\alpha$ false-alarm control under arbitrary within-episode stopping times.
   \item \textbf{Finite-horizon boundary shaping:} Inspired by finite horizon safety verification, we introduce a deterministic time-varying scaling of the monitoring boundary to tailor the monitor to a prescribed finite deployment horizon. By adjusting a temporal scaling parameter, the resulting boundary can prioritize either conservative early monitoring ($a\geq1$) or more responsive early detection ($0<a<1$). We establish a weighted false-alarm budget and derive stepwise cumulative false-alarm bounds for every $a>0$, precisely characterizing the trade-off between early detection and false-alarm control. In particular, the proposed scaling preserves the level-$\alpha$ guarantee when $a\geq1$, while providing an explicit, generally higher false-alarm bound when $0<a<1$.
   \item \textbf{Monitoring-guided intervention and adaptation}: We connect runtime barrier auditing with alarm-triggered intervention and data-driven updates of barriers and deployed policies. Using the online anytime-Valid runtime monitoring, it not only provides statistical guarantees regarding false positives but also offers additional intervention capabilities without pre-calibration.
\end{enumerate}

\section{Related Work}
\noindent\textbf{1). Runtime Verification and Statistical Monitoring.} Runtime Verification (RV) monitors executions against formal specifications such as linear temporal logic (LTL) or signal temporal logic (STL) \cite{bauer2011runtime}, but struggles in stochastic environments. Statistical Model Checking (SMC) \cite{clarke2011statistical} addresses stochastic systems through fixed-horizon sampling, but lacks time-uniform guarantees and support for arbitrary stopping times. Recent predictive RV approaches combine conformal prediction with STL specifications \cite{lindemann2023conformal,zhao2024robust,cairoli2023conformal} and adaptive conformal inference \cite{gibbs2021adaptive} to maintain coverage under distribution shifts. However, these methods provide only marginal or asymptotic guarantees and cannot ensure anytime-valid false-alarm control. Rao et al.~\cite{rao2026geometryflowmatchinguncertaintycostfree} detect VLA policy failures with a CUSUM detector~\cite{page1954continuous} whose threshold is calibrated on successful rollouts. Since our monitor instead tests the barrier's drift condition and needs no calibration set, we do not include CUSUM as an experimental baseline to detect barrier function in the main text. Several works incorporate runtime monitoring into safety assurance. The adaptive shielding approach of \cite{feng2025adaptive} refines verified safe control envelopes online, but requires an expert-specified dynamics model and a verified fallback policy. Similarly, the shielding methods of \cite{alshiekh2018safe,jansen2020safe} assume a known system model (deterministic or MDP) for offline safety verification. Relatedly, \cite{yu2025neural} develops runtime monitors for detecting or predicting violations of learned certificates in deterministic continuous-time systems and uses the collected data for offline repair. In contrast, our framework targets stochastic systems, where barrier conditions are defined through conditional expectations, and formulates runtime monitoring as an anytime-valid sequential hypothesis test based on e-processes \cite{ramdas2023game,shafer2011test,vovk2021values}. The method is model-free, relying only on online observed trajectories, and provides evidence-triggered runtime intervention with uniform false-alarm control under arbitrary stopping times within each episode.

{\brokenpenalty=0 \noindent\textbf{2). Barrier Certificates and Synthesis.} Barrier certificates have been widely studied for safety verification and controller synthesis~\cite{prajna2004safety,jagtap2018temporal,cao2025comparative,xue2025finite,ames2019control,vzikelic2023learning,zhi2024unifying,abate2024stochastic,xue2024reach,henzinger2025supermartingale,XUE2026112919}. Existing data-driven methods~\cite{mathiesen2023inner,nejati2023data,salamati2024data,wu2026pac,mathiesen2026data} focus on synthesizing barrier certificates under assumptions such as known dynamics. In contrast, our work does not synthesize barrier certificates; rather, it provides an anytime-valid runtime monitor for auditing deployed barrier certificates and detecting statistical evidence that their underlying drift conditions may no longer be consistent with the observed dynamics, particularly under distribution shifts. The resulting alarm-triggered trajectories can be reused as counterexamples for subsequent certificate updating using existing synthesis methods. This creates a complementary relationship with CEGIS~\cite{abate2021fossil,zhao2023formal,chatterjee2023learner}: whereas CEGIS focuses on iterative certificate synthesis and refinement, our framework focuses on deployment-time monitoring of a learned certificate.\par}

\paragraph{Paper organization.}
The remainder of the paper is organized as follows. Section~\ref{sec:pre} presents the problem formulation. Section~\ref{sec:methods} develops the proposed anytime-valid runtime monitoring framework, including the e-process construction, adaptive betting strategy, and finite-horizon boundary shaping. Section~\ref{sec:ex} evaluates the monitor on stochastic benchmark systems under both valid/invalid barrier certificates and a VLA policy.  Section~\ref{sec:con} concludes the paper and discusses future directions.
\section{Preliminaries}
\label{sec:pre}

This section introduces the mathematical setting underlying our framework. We first review e-processes and anytime-valid inference, and then formalize the stochastic control system and the runtime safety monitoring problem.

\subsection{E-Processes and Anytime-Valid Inference}

We work on a probability space $(\Omega,\mathcal{F},\mathbb{P})$, where $\Omega$ is the sample space, $\mathcal{F}$ is a $\sigma$-algebra, and
$\mathbb{P}$ is a probability measure. Random variables are measurable functions defined on this probability space.

\begin{definition}[Filtration]
A filtration is an increasing sequence of sub-$\sigma$-algebras $\{\mathcal{F}_t\}_{t\ge 0}$ of $\mathcal{F}$ such that $\mathcal{F}_s \subseteq \mathcal{F}_t$ for all  $s \le t$.
\end{definition}
A filtration models the information available at each time $t$. $\mathcal{F}_t$ contains exactly what is known up to that point.

\begin{definition}[Martingale and supermartingale] Let $(M_t)_{t\ge 0}$ be a sequence of random variables adapted to $(\mathcal{F}_t)_{t\ge 0}$. We say that $(M_t)_{t\geq 0}$ is a \emph{martingale} if
\begin{enumerate}
    \item $M_t$ is $\mathcal{F}_t$-measurable for all $t \ge 0$,
    \item $\mathbb{E}[|M_t|] < \infty$ for all $t \ge 0$,
    \item $\mathbb{E}[M_{t+1} \mid \mathcal{F}_t] = M_t$ almost surely for all $t \ge 0$.
\end{enumerate}
If the equality $\mathbb{E}[M_{t+1} \mid \mathcal{F}_t] = M_t$ is weakened to $\mathbb{E}[M_{t+1} \mid \mathcal{F}_t] \le M_t$, then $(M_t)$ is called a \emph{supermartingale}. In this case the process drifts downward in conditional expectation. Every martingale is a supermartingale.
\end{definition}

When the filtration is understood, we simply refer to $(M_t)_{t\geq 0}$ as a martingale or supermartingale.

\begin{definition}[E-process and test supermartingale]
A nonnegative adapted process $(M_t)_{t\ge 0}$ is called an \emph{e-process} under a null hypothesis $H_0$ if $\mathbb{E}_{H_0}[M_\tau] \le 1$ for every bounded stopping time $\tau$. A \emph{test supermartingale} is an e-process that is also a supermartingale with $M_0 = 1$. Every test supermartingale is an e-process, but the converse is not true.
\end{definition}

The power of these objects comes from a time-uniform tail bound. While Markov's inequality controls a single nonnegative random variable, Ville's inequality extends the control to the whole trajectory of a test supermartingale.

\begin{proposition}[Ville's inequality \cite{ville1939etude}]
Let $(M_t)_{t\ge 0}$ be a nonnegative test supermartingale under $H_0$. Then for any $\alpha \in (0,1)$, $\mathbb{P}_{H_0} \left(\sup_{t \ge 0} M_t \ge \frac{1}{\alpha}\right) \le \alpha$.
\end{proposition}
If the process ever crosses the threshold $1/\alpha$, we may reject $H_0$ with a false-alarm probability at most $\alpha$. %

A  method to construct a test supermartingale is  \emph{e-variables}. Let $(E_t)_{t\ge 1}$ be a sequence of nonnegative random variables such that $E_t$ is $\mathcal{F}_t$-measurable and $\mathbb{E}_{H_0}[E_t \mid \mathcal{F}_{t-1}] \le 1$. Then the product process $M_t := \prod_{k=1}^t E_k$ with $M_0 := 1$ is a nonnegative supermartingale and hence an e-process. Each factor $E_k$ can be seen as a local bet against the null and the product aggregates this evidence multiplicatively. Persistent small violations of the null accumulate steadily, while isolated fluctuations have little long-term effect. By Ville's inequality, the test that stops and rejects $H_0$ the first time $M_t \ge 1/\alpha$ controls the type-I error at level $\alpha$, uniformly over all stopping times.

\subsection{Problem Formulation}
We consider a discrete-time stochastic dynamical system
\begin{equation}
\label{system}
x_{k+1} = f(x_k, u_k, \omega_k),
\end{equation}
where $x_k \in \mathbb{R}^n$ is the state, $u_k \in \mathcal{U}$ is the control input, and $\omega_k$ denotes an independent and identically distributed (i.i.d.) stochastic disturbance. \textit{Throughout this paper, both the system dynamics $f$ and the disturbance distribution are assumed to be unknown.} The system is operated under a feedback policy $\pi:\mathbb{R}^n \to \mathcal{U}$, yielding the closed-loop dynamics $u_k = \pi(x_k)$. Let $(\mathcal{F}_k)_{k\ge 0}$ denote the natural filtration generated by the trajectory.

\paragraph{Safety objective.}
Let $\mathcal{S} \subseteq \mathbb{R}^n$ denote the safe set, $\mathcal{X}_0 \subset \mathcal{S}$  the initial set. The safety objective is to ensure that trajectories starting from an initial state $x_0 \in \mathcal{X}_0$ remain within $\mathcal{S}$ with high probability, i.e.,
$\mathbb{P} \left( x_k \in \mathcal{S}, \ \forall k \ge 0 \right)$.

A standard approach to certifying such properties is through \emph{barrier certificates}. To formalize the analysis, it is convenient to introduce the stopped (or frozen) system
\[
\hat{x}_{k+1} =
\begin{cases}
f(\hat{x}_k, \pi(\hat{x}_k), \omega_k), & \text{if } \hat{x}_k \in \mathcal{S},\\
\hat{x}_k, & \text{otherwise}.
\end{cases}
\]
This construction ensures that it remains outside once the trajectory exits $\mathcal{S}$. The probability the original system stays within $\mathcal{S}$ coincides with the probability the stopped system never leaves $\mathcal{S}$ \cite{yu2023safe}.

\paragraph{Barrier certificate condition \cite{kushner1967stochastic,yu2023safe}.}
Assume that the barrier function $B:\mathbb{R}^n\to\mathbb{R}_{\ge0}$ is integrable such that the following conditional expectation is well-defined. The function $B$ is called a barrier certificate if it satisfies
\begin{subequations}
\label{eq:barrier}
\begin{align}
&B(x) \ge 0, && \forall x \in \mathbb{R}^n, \label{eq:barrier-a}\\
&B(x) \ge 1, && \forall x \notin \mathcal{S}, \label{eq:barrier-b}\\
&\mathbb{E} \left[B(\hat{x}_{k+1}) \mid \hat{x}_k = x \right] \le B(x), && \forall x \in \mathbb{R}^n. \label{eq:barrier-supermartingale}
\end{align}
\end{subequations}
The last condition states that $B$ behaves as a supermartingale along the stopped system. If such a function exists, then for any $x_0 \in \mathcal{X}_0$, $\mathbb{P} \left( x_k \in \mathcal{S}, \ \forall k \ge 0 \right) \ge 1 - B(x_0)$.

\begin{remark}
In this paper, we consider the barrier condition of the form \eqref{eq:barrier-supermartingale} for safety verification. Nevertheless, the proposed method is not restricted to this specific formulation and can be extended to more general barrier conditions, such as $\mathbb{E} \left[B(\hat{x}_{k+1}) \mid \hat{x}_k = x \right] \le a B(x)+b, \forall x \in \mathbb{R}^n$, which is widely used in finite-time and infinite-time safety verification, as well as reach-avoid analysis \cite{xue2025finite,XUE2026112919}.
\end{remark}

\paragraph{Problem of interest.}
We consider the setting where both the barrier certificate $B$ and the controller $\pi$ are \emph{learned from data}, so condition \eqref{eq:barrier-supermartingale} may fail globally despite appearing sound on the training data. Rather than assuming the correctness of these learned components, our goal is to continuously audit their validity from realized trajectories while maintaining a time-uniform false-alarm probability bounded by a prescribed level $\alpha$. The resulting monitor can serve as a component of a runtime detection mechanism, providing statistical evidence that the learned barrier certificate is no longer supported by the observed dynamics.%

\section{Anytime-Valid Runtime Monitoring of Learned Barrier Certificates}
\label{sec:methods}
\begin{figure}[!t]
    \centering
        \includegraphics[width=\linewidth]{\here 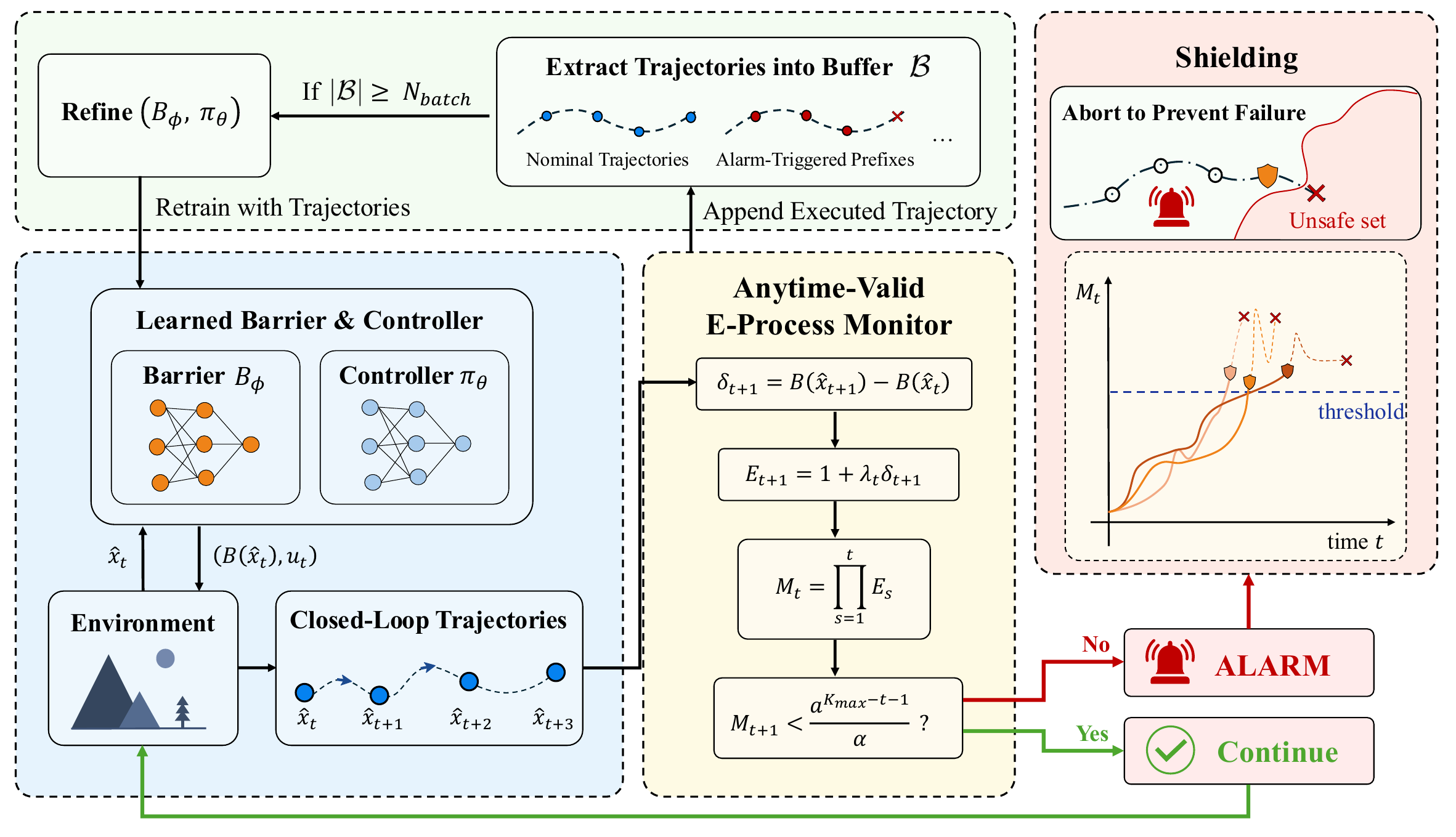} 
    \caption{Overview of the proposed framework for anytime-valid monitoring of learned barrier certificates. }
    \Description{Overview of trajectory observation, barrier evaluation, sequential evidence accumulation, and alarm-triggered refinement.}
\label{fig:framework}
\end{figure}

In this section, we introduce our anytime-valid sequential monitoring framework  for runtime auditing of learned barrier certificates (Fig. \ref{fig:framework}). 
Instead of requiring offline validation of a deployed barrier certificate, we continuously audit its consistency with observed closed-loop trajectories. By formulating barrier-condition monitoring as a sequential hypothesis test and constructing an e-process that accumulates evidence online, we obtain an anytime-valid monitor with finite-sample false-alarm control under arbitrary stopping times within an episode. The resulting evidence can be used to trigger predefined runtime interventions when the validity of the deployed safety certificate is called into question.%

\subsection{Anytime-Valid Monitoring}

\paragraph{\textbf{Trajectories and Barrier Consistency Hypothesis.}}

\emph{We first formalize the observed closed-loop trajectories and the barrier consistency hypothesis that the monitor will audit throughout deployment.}
A single run of the closed-loop system under the learned controller $\pi$ produces a trajectory $\xi  \coloneqq \{\hat{x}_0 , \hat{x}_1 , \hat{x}_2 , \dots\}$, with natural filtration $\mathcal{F}_t  \coloneqq \sigma(\hat{x}_0 ,\dots,\hat{x}_t)$, which captures all information available up to time $t$ within that specific episode, where $\hat{x}_{k+1} =f(\hat{x}_k ,\pi(\hat{x}_k ),w_k )$, $k=0,1,2,\ldots$. The first two conditions of the barrier certificate $B$ \eqref{eq:barrier-a} and \eqref{eq:barrier-b} can be constructed alone.  Our task is to test whether the conditional supermartingale condition \eqref{eq:barrier-supermartingale} remains valid under the deployed dynamics. The monitor uses one realized trajectory to accumulate statistical evidence against this conditional null.  The null hypothesis is therefore
\begin{equation}
H_0:\quad \mathbb{E}\!\Bigl[B(\hat{x}_{t+1} ) \mid \mathcal{F}_t \Bigr] \le B(\hat{x}_t ),
\label{eq:null-trajectories}
\end{equation}
for all $t \ge 0$.Define the one-step increment $\delta_{t+1}  \coloneqq B(\hat{x}_{t+1} ) - B(\hat{x}_t ),$
we test whether $\mathbb{E}[\delta_{t+1} \mid\mathcal{F}_t ]\le 0$.
Under the null the barrier shouldn't drift upward in expectation and positive increments can therefore contribute evidence against the certificate condition, with their contribution determined by the betting strategy.

\paragraph{\textbf{Local Evidence.}} We now convert each observed increment into an evidence factor that behaves like a fair bet against the null. Define $E_{t+1}  \coloneqq 1 + \lambda_t  \delta_{t+1},$ where the \emph{betting fraction} $\lambda_t \ge 0$ is chosen at time $t$ using only information available up to that point in the same episode ($\mathcal{F}_t $-measurable).  Intuitively, if the barrier increases ($\delta_{t+1} >0$), the factor exceeds $1$, signalling a surprise. If the barrier decreases, the factor is $\le 1$, consistent with $H_0$.

\begin{proposition}[One-step validity]
Under $H_0$, if $\lambda_t \ge 0$, then $\mathbb{E}_{H_0}\!\Bigl[E_{t+1}  \mid \mathcal{F}_t \Bigr] \le 1$.
\end{proposition}
\begin{proof}
By definition, $\mathbb{E}_{H_0}\!\Bigl[E_{t+1}  \mid \mathcal{F}_t \Bigr]
= 1 + \lambda_t  \mathbb{E}_{H_0}\!\Bigl[\delta_{t+1}  \mid \mathcal{F}_t \Bigr]$. Under $H_0$ the conditional expectation of $\delta_{t+1} $ is non-positive,
and $\lambda_t \ge 0$, we have the conclusion.
\end{proof}

Thus, under the null, each factor $E_{t+1} $ has conditional expectation at most $1$ -- it is a ''fair bet'' in the sense that we cannot expect to accumulate evidence against a true null.  The betting fraction $\lambda_t $ controls the sensitivity of the monitor: larger values react more strongly to positive increments but require a tighter bound to preserve non-negativity.

\paragraph{\textbf{Evidence Accumulation.}}
Local bets are combined multiplicatively into a single cumulative score that reflects the total evidence against the null accumulated during the episode. The trajectory-level evidence process is
$M_t  \coloneqq \prod_{s=1}^{t} E_{s} , M_0  \coloneqq 1.$
Because each factor $E_{s}$ has conditional expectation $\le 1$,  the product is a non-negative supermartingale under $H_0$. 

\begin{lemma}
\label{lemma1}
Under $H_0$ and $E_{s} \geq 0$, $\forall s\in \mathbb{N}$, $(M_t )_{t\ge 0}$ is a test supermartingale.
\end{lemma}
\begin{proof}
Non-negativity is obvious.  The process is adapted, and
$\mathbb{E}_{H_0}[M_{t+1} \mid\mathcal{F}_t ]
= M_t \,\mathbb{E}_{H_0}[E_{t+1} \mid\mathcal{F}_t ]
\le M_t$, where we used that $M_t $ is $\mathcal{F}_t $-measurable and the
one-step bound from Proposition~2.
\end{proof}

The product form of the e-process enables sustained evidence against the barrier condition to accumulate multiplicatively over time. When the observed barrier increments consistently provide evidence against $H_0$, the corresponding betting factors can compound the evidence substantially, whereas transient fluctuations whose effects do not persist have less lasting influence on the accumulated score. This allows the monitor to respond to systematic departures from the learned barrier condition without treating every stochastic fluctuation as a certificate violation.

\textit{Because $\{M_t\}_{t\geq 0}$ is a test supermartingale, Ville's inequality applied to the stopping rule $\tau \coloneqq \inf\{t\geq 1: M_t \ge 1/\alpha\}$ bounds the false-alarm probability of any episode by $\alpha$: if the evidence process crosses the threshold, we reject $H_0$, and under $H_0$ this occurs with probability at most $\alpha$.} The threshold crossing should therefore be interpreted as evidence that the deployed barrier condition has lost statistical support, rather than as a direct prediction of unsafe entry. The following subsection exploits a prescribed finite deployment horizon to introduce a time-varying monitoring boundary.

\subsection{Data-driven adaptive betting on \texorpdfstring{$\lambda_t$}{lambda t} (DAB)}

Even if the above theory is established, we still encounter the following two difficulties: 1. How to choose a suitable $\lambda_t$ that can satisfy the nonnegativity constraint on $E_s$ in Lemma~\ref{lemma1}; 2. Even if the constraints are met, how to adapt
these bets to the observed barrier increments? Below we will introduce $\lambda_t$ based on data-driven and use discounted empirical moments inspired by Adam.

Building on the Kelly betting principle in \cite{waudbysmith2022estimatingmeansboundedrandom,betting2021}, we use an exponentially weighted version of the empirical first and second moments, inspired by the moment estimates in Adam \cite{kingma2014adam}. The unweighted case is recovered when $\gamma=1$. Specifically, we choose $\lambda_t$ based only on barrier increments observed up to time $t$, $\delta_s=B(\hat{x}_s)-B(\hat{x}_{s-1})$ by setting $S_t=\gamma S_{t-1}+\delta_t,Q_t=\gamma Q_{t-1}+\delta_t^2$, and 
\[
\lambda_t^{\mathrm{K}}
=\max\!\left(
\frac{S_t}
{Q_t+\varepsilon},0
\right), \varepsilon>0
\]
where $\gamma\in(0,1]$ is a fixed memory parameter. Both moments contain only increments observed by time $t$, before observing
$\delta_{t+1}$. We set $S_0=Q_0= \lambda_0=0$, and
$\varepsilon=1\times10^{-12}$, and then use
\begin{equation}
\label{eq:safe-bet}
\lambda_t=
\min\!\left\{
\lambda_t^{\mathrm K},\ U,\ -\frac{1}{\underline\delta_t}
\right\}\geq0.
\end{equation}
Here, $U>0$ prevents excessively large bets and
$\underline\delta_t=-B(\hat{x}_t)$. Then
$\delta_{t+1}=B(\hat{x}_{t+1})-B(\hat{x}_t)
\geq-B(\hat{x}_t)=\underline\delta_t$, leading to the conservative cap
$\lambda_t\leq1/B(\hat{x}_t)$. Under this choice,
$E_{t+1}=1+\lambda_t\delta_{t+1}\geq1-1\geq0$.

The intuition follows a growth-rate and Kelly-type principle. When the observed barrier increments fluctuate around zero, the estimated conditional drift remains small, resulting in a conservative betting fraction. In contrast, when the observed increments provide sustained evidence of a positive barrier drift, the estimated drift increases $\lambda_t$, allowing the e-process to accumulate evidence more rapidly. Thus, the adaptive betting rule places greater weight on persistent departures from the learned barrier condition while remaining conservative when the observed dynamics provide little evidence against it. %

We introduce $\gamma$ to address a specific cancellation effect in barrier increments: without forgetting ($\gamma=1$), $S_t=\sum_{s=1}^{t}\delta_s=B(\hat{x}_t)-B(\hat{x}_0)$, so a sustained recent increase can remain masked by an earlier decrease, whereas $\gamma<1$ discounts stale increments and allows the betting fraction to respond to increases relative to a recent barrier baseline. In the appendix, we provide a detailed demonstration of the scenarios in which the $\gamma=1$ case fails.

\subsection{Shaping the Abort Time via Deterministic Scaling}
\label{sub:ds}

This subsection uses the finite deployment horizon $K_{\max}$ to shape the alarm boundary over the episode. We first rescale the evidence by the deterministic factor $a^{t}$, which turns the constant threshold $1/\alpha$ into the time-varying boundary $a^{K_{\max}-t}/\alpha$, and shows that for $a\ge1$ the probability of a false alarm within the episode is at most $\alpha$ (Proposition~\ref{thm:global}). And then extends the conclusion to every $a\geq 0$ and  refines this guarantee to every step: the probability of a false alarm by step $t$ (Theorem~\ref{thm:stepwise}).

While the constant-threshold rule remains valid under arbitrary stopping times, it does not exploit the finite deployment horizon $K_{\max}$ that often arises from practical constraints such as physical timeouts, fuel limits, or fixed simulation lengths. A finite horizon allows the monitoring boundary to be shaped so that early alarms require stronger evidence, reducing sensitivity to transient fluctuations, while the threshold gradually approaches the standard anytime-valid level as the horizon is reached. Conversely, a boundary that is lower early in the episode flags an invalid certificate sooner. We therefore introduce a deterministic temporal scaling of the boundary that adjusts the monitor's conservatism over the episode.

Inspired by finite-horizon safety verification in \cite{xue2025finite}, we introduce a deterministic time-dependent scaling of the test supermartingale to shape the monitoring boundary over a prescribed deployment horizon. Specifically, let
$M_t^a=a^tM_t,\qquad a>0.$
Under $H_0$, the supermartingale property of $(M_t)_{t\geq0}$ implies that
\[
\begin{split}
\mathbb{E}_{H_0}[M_{t+1}^a\mid\mathcal{F}_t]=a^{t+1}\mathbb{E}_{H_0}[M_{t+1}\mid\mathcal{F}_t]\leq a^{t+1}M_t
=aM_t^a.
\end{split}
\]
Thus, the conditional expectation of the scaled process grows by at most a factor of $a$ per step. For $a\geq1$, this property allows us to apply the finite-horizon martingale argument of Theorem~1 in \cite{xue2025finite}, which, together with $M_0^a=1$, yields a level-$\alpha$ bound on the probability that $M_t^a$ reaches the terminal threshold $a^{K{\max}}/\alpha$ within the prescribed horizon. This motivates the following stopping rule $\tau_a:=\inf\left\{t\in{1,\ldots,K_{\max}}:
M_t^a\geq\frac{a^{K_{\max}}}{\alpha}\right\}$,
where $\inf\emptyset=\infty$. Equivalently, substituting $M_t^a=a^tM_t$, we can express the stopping rule directly in terms of the original test supermartingale:
\[
\tau_a:=\inf\left\{t\in{1,\ldots,K_{\max}}:
M_t\geq\frac{a^{K_{\max}-t}}{\alpha}\right\}.
\]
The resulting time-dependent boundary determines how the monitoring sensitivity varies over the episode, while the underlying evidence process remains unchanged. We next establish the corresponding false-alarm guarantee. The details of the proof are provided in the Appendix~\ref{app:benchmarks}.

\begin{proposition}
\label{thm:global}
Let $H_0$ be as in \eqref{eq:null-trajectories}. Suppose that $(M_t)_{t\ge0}$ is a nonnegative test supermartingale under $H_0$. Let $a\ge1$, $K_{\max}\in\mathbb{N}$, and $\alpha\in(0,1)$. Define the event $\mathcal R\coloneqq\{\tau_a\le K_{\max}\}$. Then
$\mathbb P_{H_0}(\mathcal R)\le\alpha$.
\end{proposition}




Proposition \ref{thm:global} establishes an upper bound on the episode-wise false-alarm probability for $a\geq1$. However, it doesn't covers the regime $0<a<1$ nor characterizes how false alarms accumulate over  time steps. To provide a more comprehensive characterization, we establish a stepwise false-alarm bound that holds for all $a>0$ and every $1\leq t\leq K_{\max}$. In particular, for $a\geq1$, setting $t=K_{\max}$ recovers the guarantee established in Proposition \ref{thm:global}.

\begin{theorem}[Stepwise falsification guarantee]
\label{thm:stepwise}
Let $H_0$ be as in \eqref{eq:null-trajectories}. Suppose that $(M_t)_{t\geq0}$ is a nonnegative test supermartingale under $H_0$. Let $a>0$, $K_{\max}\in\mathbb N$ with $K_{\max}\geq1$, and $\alpha\in(0,1)$. Then the following hold:
\begin{enumerate}[label=(\roman*),leftmargin=*,nosep]
\item \emph{Weighted false-alarm budget.} The distribution
of the first alarm time satisfies
$\sum_{s=1}^{K_{\max}}
a^{K_{\max}-s}\,
\mathbb P_{H_0}(\tau_a=s)\leq\alpha.$

\item \emph{Cumulative false-alarm bound.} For every
$1\leq t\leq K_{\max}$,
\[
\mathbb P_{H_0}(\tau_a\leq t)\leq
\begin{cases}
\alpha\,a^{-(K_{\max}-t)}, & a\geq1,\\
\min\{1,\alpha\,a^{-(K_{\max}-1)}\}, & 0<a<1.
\end{cases}
\]
\end{enumerate}
over the whole horizon $\mathbb P_{H_0}(\tau_a\leq K_{\max})
\leq
\min\left\{1,\alpha\max\{1,a^{1-K_{\max}}\}\right\}$.
\end{theorem}


\begin{remark}[Tightness and limits]
\label{rmk:tight}
The bound (ii) is tight. For $a\geq1$ and any fixed step $t$, consider the martingale that remains at $1$ until step $t$, when it jumps to $a^{K_{\max}-t}/\alpha$ with probability $\alpha a^{-(K_{\max}-t)}$ and to $0$ otherwise, and remains constant thereafter. This martingale attains the bound, since $\mathbb P_{H_0}(\tau_a\leq t)=\alpha a^{-(K_{\max}-t)}$. The same construction applies to any deterministic boundary $h$: if $h(t)<1/\alpha$ at some step $t$, the jump at that step raises an alarm with probability $\min\{1,1/h(t)\}>\alpha$. A boundary therefore keeps level $\alpha$ for every evidence process only if it stays at or above $1/\alpha$. If level $\alpha$ and early rejection are both required, deterministic scaling cannot deliver them. 
\end{remark}


Ultimately, this sample scaling mechanism has profound consequences for the temporal dynamics of the abort time. 

\begin{itemize}
\item \textbf{Case $a = 1$:} $M^a_t = M_t$ is a test supermartingale; the rejection threshold is $1/\alpha$ and the test is valid for any stopping time (no bias toward early or late abort). While theoretically clean, it ignores the finite horizon and treats all time steps equally, which can be wasteful when resources are limited or early decisions are preferred.

\item \textbf{Case $a > 1$ (patient test):} $a^{K_{\max}-t}$ decreases exponentially over time, making the rejection threshold $\frac{a^{K_{\max}-t}}{\alpha}$ highest initially and lowest near the deadline. Thus, early alarms require stronger evidence, while the test approaches the classical anytime-valid rule as $t\rightarrow K_{\max}$. This makes rejection easier as the experiment proceeds, thus favouring \textbf{later abort}: early evidence must be overwhelming to trigger an alarm, while near the deadline the bar is no higher than the classical anytime-valid test. For example, a drone may experience short-lived oscillations after takeoff. A patient monitor avoids unnecessary aborts by requiring sustained evidence before intervention. Larger values of $a$ improve robustness against transient fluctuations and spurious alarms, while values closer to $1$ recover the standard monitor. This conservatism, however, reduces sensitivity to brief or rapidly developing safety violations.

\item \textbf{Case $a < 1$ (impatient test):} The factor $a^{K_{\max}-t}$ increases exponentially with time. Consequently, the monitoring threshold $\frac{a^{K_{\max}-t}}{\alpha}$ is lowest at the beginning of the episode and increases toward the deadline. This makes the monitor more sensitive to early evidence: relatively modest evidence can trigger an alarm at the beginning, whereas progressively stronger evidence is required later. Such a strategy may be desirable in safety-critical applications where delayed intervention is costly. For example, an autonomous vehicle approaching a pedestrian crossing may benefit from an early alarm when observations indicate that the deployed barrier condition no longer reflects the current environment. A smaller value of $a$ further lowers the initial threshold, increasing early sensitivity, while values closer to $1$ yield a less pronounced early-warning bias. However, this increased sensitivity comes at the cost of weaker false-alarm control. In particular, the standard level-$\alpha$ guarantee no longer holds. Instead, the whole-horizon false-alarm probability is bounded by $\min\{1,\alpha a^{-(K_{\max}-1)}\}$. Moreover, the increasing threshold makes the monitor less sensitive to evidence arising later in the episode. Thus, choosing $a<1$ favors early intervention at the expense of a potentially higher false-alarm probability and reduced sensitivity to slowly emerging violations.

\end{itemize}

By framing $a$ as a patience parameter, the safety engineer can deliberately prioritize robustness against early noise or rapid detection and resource efficiency, choosing the mode that best aligns with the system's physical characteristics and the cost of data collection.



\subsection{Comparison with Direct Barrier Monitoring}
\label{sub:direct}

Since the barrier value $B(\hat{x}_t)$ is a supermartingale under $H_0$, one may ask whether directly monitoring the normalized barrier process $B(\hat{x}_t)/B(\hat{x}_0)$ is sufficient, making the proposed DAB mechanism unnecessary. Indeed, Ville's inequality guarantees episode-wise false-alarm control at level $\alpha$ by raising an alarm whenever $\frac{B(\hat{x}_t)}{B(\hat{x}_0)}\geq \frac{1}{\alpha}$. Thus, direct barrier monitoring is statistically valid under $H_0$. Moreover, as shown in Remark~\ref{rmk:ratio}, it is a special case of the proposed multiplicative e-process. The key distinction is that direct monitoring fixes the betting strategy, whereas DAB allows the betting strategy to adapt to the observed barrier increments.

\begin{enumerate}
    \item \textbf{Dependence on the initial barrier value.}
    Both direct and DAB monitoring use the alarm threshold $1/\alpha$ when their evidence processes are normalized to start at $1$. However, direct monitoring is based on the relative barrier value $B(\hat{x}_t)/B(\hat{x}_0)$, so its equivalent threshold in terms of the unnormalized barrier is $B(\hat{x}_0)/\alpha$. Consequently, the response of the normalized barrier process to a given absolute barrier change depends on the initial barrier value. In contrast, DAB provides greater flexibility in how barrier increments contribute to the monitoring statistic through its adaptive betting strategy.

    \item \textbf{Adaptive evidence accumulation.}
    Direct monitoring corresponds to the particular betting strategy $\lambda_t=1/B(\hat{x}_t)$, for which the multiplicative e-process reduces to $M_t=\frac{B(\hat{x}_t)}{B(\hat{x}_0)}$. DAB allows alternative adaptive betting strategies that respond to recent barrier increments. In our implementation, DAB refrains from betting when the discounted sum $S_t$ of recent increments is negative, avoiding evidence losses during these periods and preserving evidence for subsequent barrier increases. As demonstrated in our experiments, this flexibility can improve detection across a broader range of initial states. Such improvements are empirical, however, and do not imply that DAB always detects certificate degradation earlier than direct monitoring.
    \item  \textbf{Patient and impatient testing.}
    The parameter $a$ controls the temporal profile of the alarm boundary $h_a(t)=a^{K_{\max}-t}/\alpha$, independently of the betting strategy. For $a>1$, the boundary is higher at earlier time steps and decreases
    to $1/\alpha$ at the end of the deployment horizon, resulting in more patient monitoring. In this case, the cumulative false-alarm probability is bounded strictly below $\alpha$ at every intermediate time step,
    while the episode-wise false-alarm probability over the entire horizon remains bounded by $\alpha$. Conversely, for $0<a<1$, the boundary is lower at earlier time steps and increases over time, resulting in more impatient monitoring. This comes at the cost of a potentially larger
    episode-wise false-alarm bound. The case $a=1$ recovers the conventional constant boundary. The corresponding trade-off is quantified by Theorem \ref{thm:stepwise}.
\end{enumerate}

\begin{remark}
\label{rmk:ratio}
Direct barrier monitoring is recovered by choosing $\lambda_t=1/B(\hat{x}_t)$, provided $B(\hat{x}_t)>0$. Specifically, with $\delta_{t+1}=B(\hat{x}_{t+1})-B(\hat{x}_t)$, $E_{t+1}
=1+\lambda_t\delta_{t+1}
=\frac{B(\hat{x}_{t+1})}{B(\hat{x}_t)}$.
Consequently, $M_t=\prod_{s=1}^{t}E_s
=\frac{B(\hat{x}_t)}{B(\hat{x}_0)}$ with $M_0=1$. Thus, direct barrier monitoring is a special case of the proposed multiplicative e-process. The potential benefit of DAB comes from its more flexible betting strategy and the resulting evidence accumulation, rather than from a stronger false-alarm guarantee.
\end{remark}

\subsection{Runtime Monitoring}
The preceding subsections establish the statistical foundations of the proposed anytime-valid monitor. We now instantiate these results as a deployment-time runtime monitoring procedure.

In each deployment episode, the learned barrier $B_\phi$ and policy $\pi_\theta$ are fixed, and the monitor updates the e-process online using the observed closed-loop trajectory. The episode is terminated when the evidence process crosses the time-varying decision boundary. Under the null hypothesis that the deployed barrier satisfies the conditional drift condition, the probability of such an alarm within an episode is controlled at the prescribed level $\alpha$, regardless of the stopping time. The monitoring procedure is independent of how the barrier is constructed and can therefore be applied to analytical, optimization-based, or learned barrier certificates, provided that the assumptions required by the monitor are satisfied. The collected trajectories can optionally be stored as counterexamples and used for subsequent barrier or policy refinement between episodes. Such updates do not occur within an episode and are therefore separate from the anytime-valid monitoring procedure. The complete deployment procedure is summarized in Algorithm \ref{alg1}.

\begin{algorithm}[!htbp]
\caption{Anytime-Valid Runtime Monitoring}
\label{alg1}
\begin{algorithmic}[1]
\REQUIRE Initial barrier $B(\hat{x}_0)$, policy $\pi_0$, buffer size
$N_{\mathrm{batch}}$, significance level $\alpha$, maximum
horizon $K_{\max}$, and patience factor $a>0$.
\ENSURE Runtime monitoring
\STATE Initialize $B_\phi\leftarrow B_0$,
$\pi_\theta\leftarrow\pi_0$, $\mathcal{B}\leftarrow\emptyset$
\FOR{deployment episode $r=1,2,\ldots$}
    \STATE Observe $\hat{x}_0$; initialize
    $M_0\leftarrow1$ and $\xi\leftarrow(\hat{x}_0)$
    \FOR{$t=0,1,\ldots,K_{\max}-1$}
        \STATE Choose  $\lambda_t\geq0$ as in
        Eq.~\eqref{eq:safe-bet}
        \STATE Set $u_t\leftarrow\pi_\theta(\hat{x}_t)$ and observe $\hat{x}_{t+1}$
        \STATE Append $(\hat{x}_t,u_t,\hat{x}_{t+1})$ to
        $\xi$
        \STATE Compute
        $\delta_{t+1}\leftarrow
        B_\phi(\hat{x}_{t+1})-B_\phi(\hat{x}_t)$
        \STATE Update
        $M_{t+1}\leftarrow M_t(1+\lambda_t\delta_{t+1})$
        \IF{$M_{t+1}\geq a^{K_{\max}-(t+1)}/\alpha$}
            \STATE Exit the episode; \textbf{break}
        \ENDIF
        \IF{$\hat x_{t+1}\notin\mathcal S$}
            \STATE  $\hat x_{t+1}$ is fixed ;  $t\leftarrow t+1 $
        \ENDIF
    \ENDFOR
    \STATE $\mathcal{B}\leftarrow\mathcal{B}\cup\{\xi\}$
    \IF{$|\mathcal{B}|\geq N_{\mathrm{batch}}$}
        \STATE \textbf{Optional:} Update $B_\phi,\pi_\theta$
        using $\mathcal{B}$; set $\mathcal{B}\leftarrow\emptyset$
    \ENDIF
\ENDFOR
\end{algorithmic}
\end{algorithm}

\begin{remark}
\label{outpoint}
The monitoring procedure is intended to operate while the system remains in the safe set $\mathcal{S}$. By construction, the stopped process follows the original system's dynamics whenever it is in the safe set $\mathcal{S}$ and remains frozen after exiting $\mathcal{S}$. Consequently, the global supermartingale condition 
$\mathbb{E}\left[
B(\hat{\bm{x}}_{k+1})
\mid \hat{\bm{x}}_k=\bm{x}
\right]
\le B(\bm{x}),
 \forall \bm{x}\in\mathbb{R}^n,$
is equivalent to the barrier-drift condition for the original system restricted to the safe set:
$\mathbb{E}\left[
B(\bm{x}_{k+1})
\mid \bm{x}_k=\bm{x}
\right]
\le B(\bm{x}),
\forall \bm{x}\in\mathcal{S}.$
Indeed, outside $\mathcal{S}$, the stopped process remains at its current state, so the supermartingale condition holds automatically with equality. Thus, the safe set $\mathcal{S}$ is precisely the region in which the barrier-drift condition needs to be monitored.  Accordingly, each trajectory is terminated upon its first exit from $\mathcal{S}$ or upon reaching the prescribed horizon $K_{\max}$, whichever occurs first. Thus, once a trajectory exits the safe set, there is no need to simulate its subsequent evolution.

\end{remark}

\begin{remark}
The monitor is designed to detect statistical evidence against the barrier condition, rather than to predict unsafe entry. At the same time, the monitor can serve as a component of a runtime safety-shielding mechanism by triggering a predefined conservative intervention when sufficient evidence indicates that the learned certificate is no longer consistent with the observed dynamics. Thus, a trajectory may move toward the unsafe region without triggering the monitor, provided that the monitored barrier condition remains statistically consistent with the observed dynamics. Such a situation does not constitute a failure of the proposed detection objective, since the monitor is intended to identify when the statistical support for the safety certificate has become questionable, rather than to predict every impending safety violation. Once sufficient evidence against the barrier condition accumulates, the deployed certificate is treated as unsupported for continued operation, and the predefined intervention can be activated to reduce the risk of subsequent unsafe behavior. 
\end{remark}

Alg. \ref{alg1} resets the e-process at the start of each episode, yielding false-alarm control within each episode. The per-episode guarantee is appropriate when episodes are treated as separate operational units, as in episodic RL or session-based monitoring. If error control is instead required across $N_{\mathrm{ep}}$ deployment episodes, the episode-wise error budget can be allocated such that $\sum_{r=1}^{N_{\mathrm{ep}}} \alpha_r\leq \alpha$, for example, using $\alpha_r=\frac{\alpha}{N_{\mathrm{ep}}}$ or a summable sequence such as $\alpha_r=\frac{\alpha}{2^{r}}$. By the union bound, the probability of at least one false alarm over all $N_{\mathrm{ep}}$ episodes is then at most $\alpha$. More elaborate cross-episode e-process constructions may also be considered when their required measurability and null conditions are satisfied. Such global error control involves a sensitivity trade-off: decreasing $\alpha_r$ across episodes increases the rejection threshold $\frac{1}{\alpha_r}$ and may reduce detection power in later episodes. We therefore use per-episode error control in our experiments and leave cross-episode error allocation as an application-dependent extension.

\section{Experiments}
\label{sec:ex}
We evaluate the anytime-valid runtime monitoring framework on three systems and on a learned vision-language-action (VLA) policy in closed loop. The experiments address four questions: 
\begin{enumerate}
    \item \textbf{RQ1:} Does the monitor preserve the prescribed false-alarm guarantee when the barrier certificate is valid?
    \item \textbf{RQ2:} Does this monitor provide evidence beyond the false-alarm bounds when the barrier certificate is invalid?
    \item \textbf{RQ3:} How does temporal boundary shaping trade the trade-off between avoiding unnecessary alarms and detecting de\-gradation promptly? 
    \item \textbf{RQ4:} What does the monitor provide in practice and intervention on a VLA policy?
\end{enumerate}

The default setting is $K_{\max}=400$, $K_{\max}\ln a=4$, $\gamma=0.92$, and $U=100$. Table~\ref{tab:validity} also reports the standard anytime-valid rule $K_{\max}\ln a=0$ ($a=1$) on the same runs, and the impatient boundary $K_{\max}\ln a=-1$. Subsection~\ref{sec:barrier} answers RQ1, Subsection~\ref{sec:invalidbarrier} answers RQ2, Subsection~\ref{exp:patience} answers RQ3, and Subsection~\ref{sec:practical} answers RQ4. Detailed benchmark descriptions, implementation details, proofs of validity or counterexamples establishing invalidity for the candidate barriers are provided in Appendix~\ref{app:benchmarks}. The sensitivity of $K_{\max}$ and $\lambda_t$ is provided in Appendix~\ref{app:baselines-full}. 

The formal objective of the proposed framework is runtime monitoring of the learned barrier condition. Theorems~\ref{thm:global} and~\ref{thm:stepwise} are the statistical guarantee of the monitor: under $H_0$, the probability of an alarm within the episode is at most $\alpha$ for $a\ge1$, and by step $t$ at most $\alpha a^{-(K_{\max}-t)}$. For $a<1$, Theorem~\ref{thm:stepwise} states the larger bound. The monitor can serve as a component of a runtime safety-shielding mechanism: when sufficient evidence is accumulated against the deployed barrier, a predefined intervention can be triggered rather than continuing to rely on an unsupported barrier. The same alarm provides an evidence-based trigger for runtime intervention, which we evaluate in Subsection~\ref{sec:practical}.
\begin{ResultBox}
\noindent\textbf{Why do we use $K_{\max}\ln a$ ?} Since $a$ enters the alarm threshold exponentially over the horizon, varying $a$ directly makes it difficult to interpret the strength of threshold shaping. We therefore parameterize $a$ in terms of $K_{\max}\ln a$, which directly determines the logarithm of the threshold inflation factor at the beginning of the horizon relative to the terminal threshold. \par
With this parameterization, the alarm
threshold at step $k$ is
\[\ln M_k \;\geq\; \ln(1/\alpha) + K_{\max}\ln a\,\Bigl(1-\frac{k}{K_{\max}}\Bigr).\]
Thus, at $k=0$, the threshold lies $K_{\max}\ln a$ above $\ln(1/\alpha)$, and it decreases linearly in $k$ to $\ln(1/\alpha)$ at $k=K_{\max}$. \par
The threshold $a^{K_{\max}-k}/\alpha$ on $M_k$ starts at $e^{K_{\max}\ln a}/\alpha$ and decreases exponentially to $1/\alpha$. For $K_{\max}\ln a<0$ the threshold starts below $\ln(1/\alpha)$ and increases to the same endpoint.
\end{ResultBox}

\subsection{Runtime Monitoring under Valid Certificates}
\label{sec:barrier}



Under a valid certificate, Theorem~\ref{thm:stepwise} guarantees that the probability of an alarm within the horizon $K_{\max}$. To empirically assess the consistency of the observed alarm behavior within this theoretical guarantee, we sample $N$ initial states $x_0^{(1)},\dots,x_0^{(N)}$ and simulate $R$ independent trajectories from each initial state. Let $\tau^{(i,j)}$ be the alarm time of the monitor on the $j$-th trajectory from $x_0^{(i)}$. The empirical alarm rate for $x_0^{(i)}$ until time $t\in[1,K_{\max}]$ is
$\hat{p}_i(t)=\frac{1}{R}\sum_{j=1}^{R}\mathbb{I}\big(\tau^{(i,j)}\le t\big),$
which provide an empirical estimate of the probability $\mathbb{P}_{H_0}(\tau_a\le t)$ in Theorem~\ref{thm:stepwise}. We also report the worst case to show the robustness: $\hat{p}^{\max}(t)=\max_{i}\hat{p}_i(t)$. To summarize the empirical alarm behavior across the sampled initial states, 
we classify $x_0^{(i)}$ as empirically abnormal at time $t$ if
\[
\hat{p}_i(t)> \text{Bound}_i(t)=
\begin{cases}
\alpha a^{t-K{\max}}, & a\geq 1,\\
\min\{1,\alpha a^{1-K_{\max}}\}, & 0<a<1,
\end{cases}
\]
and report the fraction of empirically abnormal initial states,
$\rho_\alpha(t)=\frac{1}{N}\sum_{i=1}^{N}\mathbb{I}\big(\hat{p}_i(t)>\text{Bound}_i(t)\big).$

Under a valid certificate and any time $t$, we expect $\rho_\alpha(t)$ to be close to zero and $\hat{p}^{\max}(t)$ to remain close to or below the bound in Theorem~\ref{thm:stepwise}, providing empirical evidence consistent with the theoretical guarantee. Conversely, observing empirical alarm rates below the bound does not establish the validity of the certificate by itself which is discussed in Appendix~\ref{sec:exp-a}.

\begin{table}[!htbp]
\caption{Alarm rates (\%) under valid and invalid certificates on the same runs ($N=R=10^3$, $K_{\max}=400$, $\alpha=5\%$) at $t=K_{\max}$. $^\dagger$Level $\alpha a^{1-K_{\max}}=0.136$ (Theorem~\ref{thm:stepwise}). }
\label{tab:validity}
\centering
\small
\setlength{\tabcolsep}{3pt}
\renewcommand{\arraystretch}{1.1}
\begin{tabular*}{\linewidth}{@{\extracolsep{\fill}} l c rr rr @{}}
\toprule
& & \multicolumn{2}{c}{Direct} & \multicolumn{2}{c}{DAB} \\
\cmidrule(lr){3-4} \cmidrule(lr){5-6}
System & $K_{\max}\ln a$ & $\rho_\alpha$ & $\hat{p}^{\max}$ & $\rho_\alpha$ & $\hat{p}^{\max}$ \\
\midrule
\multicolumn{6}{l}{\textit{Valid certificates}} \\
System~1   & $-1^\dagger$ & 0.00 & 11.40 & 0.00 & 3.50 \\
           & 0 & 0.00 & 4.80 & 0.00 & 0.90 \\
           & 4 & 0.00 & 0.20 & 0.00 & 0.10 \\
System~2   & $-1^\dagger$/0/4 & 0.00 & 0.00 & 0.00 & 0.00 \\

\midrule
\multicolumn{6}{l}{\textit{Invalid certificates}} \\
System~2-input  & $-1^\dagger$ & 18.10 & 100.00& 82.60 & 99.20 \\
               & 0 & 18.10 & 100.00 & 82.10 & 92.00 \\
               & 4 & 17.90 & 100.00 & 79.90 & 81.50 \\
System~3       & $-1^\dagger$ & 28.40 & 69.40 & 80.60 & 99.80 \\
               & 0 & 24.00 & 69.40 & 79.00 & 96.30 \\
               & 4 & 23.80 & 69.40 & 78.40 & 95.10 \\

\bottomrule
\end{tabular*}
\end{table}

\noindent \textbf{Results. }We evaluate the monitor on two benchmark systems for which globally valid barrier certificates satisfying \eqref{eq:barrier} are available. We compare two betting strategies: DAB and Direct, which uses $\lambda_t=1/B(\hat{x}_t)$. As shown in Table~\ref{tab:validity}, for $K_{\max}\ln a\in\{0,4\}$, both monitors achieve $\hat{p}^{\max}(K_{\max})<\alpha$, empirically consistent with the global false-alarm guarantee $\mathbb{P}_{H_0}(\tau_a\le K_{\max})\le\alpha$ in Proposition~\ref{thm:global}. Under the more permissive boundary with $K_{\max}\ln a=-1$, the theoretical upper bound is $13.6\%$. The observed maximum empirical alarm rates are $11.40\%$ for Direct and $3.50\%$ for DAB. Across the tested settings, Direct generally exhibits higher empirical alarm rates than DAB. This difference is particularly evident in System~1, where the initial barrier values $B(x_0)$ are relatively small, as illustrated in Figure~\ref{fig:alarmrate2}. For Direct, a small initial value makes the alarm condition $B(\hat{x}_t)/B(\hat{x}_0)\geq 1/\alpha$ more sensitive to subsequent increases in the barrier value. In contrast, DAB adapts its betting strategy based on the observed barrier increments and is less sensitive to the initial barrier value.
Figure~\ref{fig:temporal-bound} further examines the stepwise false-alarm guarantee in Proposition~\ref{thm:stepwise}. For System~1, the maximum empirical alarm rate across the sampled initial states remains below the corresponding theoretical bound at each time step, although the empirical rate exhibits some transient increases during the early stages. For System~2, DAB produces no alarms over the simulated trajectories. Therefore, we omit its results from the figure. Overall, these empirical observations are consistent with the global and stepwise false-alarm guarantees under valid barrier certificates.

\subsection{Runtime Monitoring under Invalid Certificates}
\label{sec:invalidbarrier}

In this subsection, we will answer RQ2 by investigating scenarios involving invalid barriers. The remaining systems in Table \ref{tab:validity} have invalid barrier certificates. We consider two scenarios: (1) initially, we obtain a valid barrier function for System~2 in RQ1. However, environmental changes are inevitable in real-world scenarios. To simulate such a scenario, we actively injected noise  $b_k\sim\mathcal U[0,0.45]$ into state $x_2$ from step $k=100$ onward:
\begin{equation}
    \begin{aligned}
        x_{2,k+1}=x_{2,k}&+0.01\big((0.8+\xi_k)\,x_{1,k}\\
        &+10\,(x_{1,k}^2-0.21)\,x_{2,k}+b_k\big)
    \end{aligned}
\end{equation}
where the equation of $x_1$ is unchanged. Consequently, the original barrier function proved to be invalid in the changed environment.  (2) The environment doesn't change, but due to the sample-based training, the barrier function for System~3 is invalid from the start. Although our monitor can deal with the black-box systems, we test the candidate barrier function against the dynamic system and use SMT to find a definite counterexample, thus proving that the candidate is indeed invalid for System~2+input and System~3.

\noindent \textbf{Results.} When $K_{\max}\ln a =4$, under invalid certificates, both methods detect substantial deviations from the prescribed alarm behavior, but their performance differs across the two systems. In System~2-input attack, direct betting has a worst-case alarm rate of $100\%$, compared with $81.5\%$ for DAB. However, DAB identifies a much larger fraction of abnormal initial states, with $\rho_{\alpha}=79.9\%$ compared with $17.9\%$ for direct betting. In System 3, DAB performs more sensitive in terms of both metrics: its $\rho_{\alpha}$ reaches $78.40\%$, compared with $23.80\%$ for direct betting, while its worst-case alarm rate is $95.10\%$, compared with $69.40\%$ for direct betting. These results highlight the complementary roles of the two metrics. A high worst-case alarm rate $\hat{p}^{\max}$ indicates that the monitor frequently raises alarms for at least one sampled initial state, suggesting that the deployed barrier condition may no longer be trustworthy. Meanwhile, $\rho_{\alpha}$ quantifies the fraction of sampled initial states for which the empirical alarm rate exceeds $\alpha$, revealing how widely this evidence of potential barrier invalidity occurs across the initial states.

\subsection{The patience parameter $a$}

\label{exp:patience}

\begin{figure*}[!t]
\centering
\includegraphics[width=0.75\textwidth]{\here 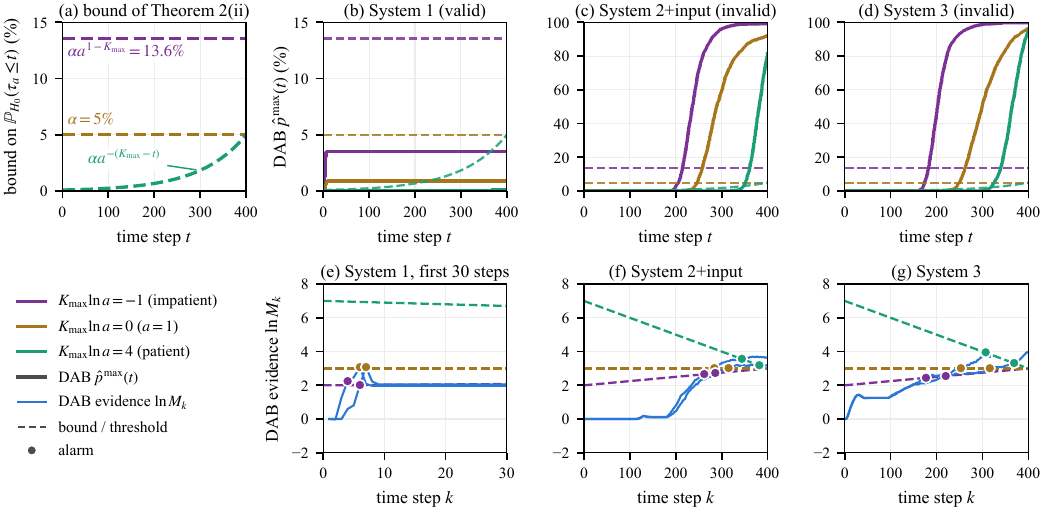}
\caption{Stepwise bound and DAB evidence for $K_{\max}\ln a\in\{-1,0,4\}$. (a) The bound of Theorem~\ref{thm:stepwise}(ii). (b--d) $\hat p^{\max}(t)$ on the runs of Table~\ref{tab:validity}: below its bound under the valid certificate, above it under the invalid ones. (e--g) Evidence of two example runs against the three thresholds.}
\Description{Two rows of panels, one column per system. Top: the false-alarm bound over time and the worst-state alarm rate of System 1, System 2+input and System 3 for three boundary settings. Bottom: example DAB evidence paths against the three thresholds, with alarms marked.}
\label{fig:temporal-bound}\label{fig:trade-off-a}
\end{figure*}
We introduce $a$ as a deployment-time design parameter for the monitoring boundary. For fixed $\alpha$, $K_{\max}$, and betting strategy, changing $a$ leaves the underlying evidence process $M_k$ unchanged and modifies only the decision rule applied to that evidence. Its choice expresses the trade-off between avoiding unnecessary intervention and detecting certificate invalidity promptly. A larger \mbox{$a$} is appropriate when unnecessary early interventions are costly and the application can tolerate delayed detection. Figure~\ref{fig:temporal-bound} shows the bound over time and $\hat p^{\max}(t)$ for DAB with $K_{\max}\ln a\in\{-1,0,4\}$. 

The results in Subsections~\mbox{\ref{sec:barrier}} and \ref{sec:invalidbarrier}, reported in Table~\mbox{\ref{tab:validity}} and Figure~\mbox{\ref{fig:alarmrate2}}, show that increasing \mbox{$a$} cannot increase the number of alarms and cannot advance the first alarm time. This observation is consistent with the theoretical analysis: a larger \mbox{$a$} raises the rejection threshold at early time steps, making the monitor conservative. To illustrate how this affects monitoring under valid and invalid certificates and also stepwise falsification gurantee at any time $t$, we consider Systems~1, System~2+input and System~3.

In general, under the valid condition, the alarm rate remains below the upper bound specified in Theorem~\ref{thm:stepwise}, regardless of the parameters of $a$. Under the invalid condition, the alarm rate will exceed this upper bound after a certain period of time. Specifically, in System~1, where the barrier certificate is valid, the DAB evidence process exhibits a transient increase that triggers an alarm when $a=1$, but then drops rapidly (we give an example trajectory in Figure~\ref{fig:trade-off-a}). Since the certificate is valid in this scenario, the threshold crossing is a false alarm caused by a temporary fluctuation, with no sustained evidence against the barrier condition. Increasing the temporal scaling to $K_{\max}\ln a=4$ raises the early-time rejection threshold, preventing the evidence process from triggering an alarm. Decreasing to $K_{\max}\ln a=-1$, these trajectories will trigger an alert. In this case, the more conservative threshold helps avoid an unnecessary alarm. In contrast, in System~2+input and System~3, where the barrier certificate is invalid, increasing the temporal scaling to $K_{\max}\ln a=4$ delays the alarm compared with the case $K_{\max}\ln a=0 \; \text{and} \; -1$, but the evidence eventually becomes sufficient to cross the rejection threshold. Also, according to Figure~\ref{fig:temporal-bound} for large a in System~2+input and System~3, increasing $a$ causes the overall false-alarm rate to exceed its stepwise falsification guarantee more quickly. Overall, the monitoring mechanism can still detect evidence that contradicts the "barrier conditions," even though such evidence may be accumulated at a later stage and has difference between different $a$. These examples illustrate the trade-off controlled by $a$: a more conservative threshold can suppress alarms caused by transient fluctuations under a valid certificate, but may also delay the detection of an invalid certificate.

In practice, the validity of a learned barrier certificate is generally unknown before deployment. Thus, the choice of $a$ should reflect the expected reliability of the certificate and the trade-off between avoiding unnecessary alarms and detecting potential degradation promptly. Setting $a=1$ provides a natural default when prompt detection is important while retaining the episode-wise level-$\alpha$ false-alarm guarantee. For a more conservative monitoring strategy, one may choose $a>1$, which raises the alarm boundary at earlier time steps. This choice is particularly suitable when transient or benign disturbances are expected during the initial phase of deployment and avoiding unnecessary alarms is a priority. Conversely, choosing $0<a<1$ lowers the boundary at earlier time steps, favoring more prompt detection at the cost of a potentially larger episode-wise false-alarm probability. This more aggressive setting is appropriate when the cost of unnecessary alarms is low and early detection of potential safety degradation takes priority.


\subsection{Monitoring-Guided Intervention and Policy Adaptation}
\label{sec:practical}

\begin{figure}[!t]
    \centering
    \includegraphics[width=\linewidth]{\here 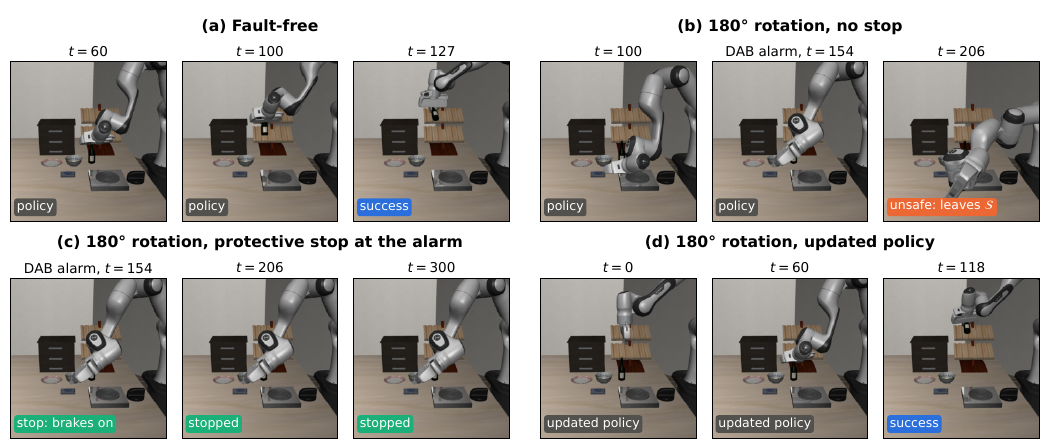}
    \caption{One VLA rollout (wine bottle on rack): (a) fault-free; (b) $180^\circ$ rotation, no stop; (c) the same fault with the protective stop at the DAB alarm; (d) the updated policy.}
    \label{fig:libero-frames}
\end{figure}

Once deployed, the monitor audits the barrier certificate from observed trajectories: it needs no model of the dynamics, no
predictive model, and only needs one calculate of the deployed barrier per step (even 0.42 ms per step for barrier evaluation on the VLA task). The alarm therefore can be used in three ways. (1) The first is the audit itself, the statistical evidence against the deployed certificate studied in Subsections \ref{sec:barrier}--\ref{exp:patience}. (2) The second is intervention: an alarm implies that the previous barrier function may no longer be applicable and engineers can choose to terminate the trajectory early. (3) The third is repair: the running trajectories are also the data for the update step of Algorithm~\ref{alg1}. We study intervention and repair on System 2+input, System 3 and the VLA. Taken together, these three elements constitute the core of a Monitoring-Guided intervention and policy adaptation.

The policy on the VLA is a pre-trained $\pi_{0.5}$ from \cite{intelligence2025pi} performing three tasks (open the middle drawer, put the bowl on the plate, put the wine bottle on the rack). We define a state to be unsafe if it leaves a workspace box $\mathcal S$. We inject six faults: the planar command is rotated by $90^\circ$, $137^\circ$ or $180^\circ$; the $90^\circ$ rotation with an actuator dead zone, so that command components below 0.05 produce no motion; or a constant force of 10 or 25\,N pushes the hand. A missed exit into the unsafe is costly on a robot, so we use the impatient boundary $K_{\max}\ln a=-1$ with $K_{\max}=300$ steps and $\alpha=0.1$ (Subsection~\ref{exp:patience}). On the VLA, once the online monitor alarms, the robot triggers its protective stop: the motion command is set to zero and the joint brakes lock the arm. Table~\ref{tab:update-effects} reports three settings from the same ten initial states per task in VLA: the policy alone (\emph{No stop}), the policy stopped at the alarm (\emph{Stop}), and the updated policy. System~2+input and System~3's settings are same as before. Since our primary focus here is the overall effect of the intervention, we report the number of alarms and unsafe trajectories across all $N \times R$ trajectories.

\begin{table}[!htbp]
\caption{Effect of the alarm (DAB), in \% of trajectories: $10^6$ per benchmark, 180 in VLA. Gray: change from No stop.}
\label{tab:update-effects}
\centering
\setlength{\tabcolsep}{4pt}
\begin{tabular*}{\columnwidth}{@{\extracolsep{\fill}}lccc}
\toprule
 & No stop & Stop & Updated \\
\midrule
\multicolumn{4}{l}{\emph{System~2+input}} \\
\quad Alarm rate & 47.27 & 47.27 & 0.00\dd{$-47.27$} \\
\quad Unsafe & 0.00 & 0.00 & 0.00 \\
\addlinespace
\multicolumn{4}{l}{\emph{System~3}} \\
\quad Alarm rate  & 46.69 & 46.69 & 0.00\dd{$-46.69$} \\
\quad Unsafe & 42.88 & 22.69\dd{$-20.19$} & 0.04\dd{$-42.84$} \\
\addlinespace
\multicolumn{4}{l}{\emph{VLA}} \\
\quad Alarm rate  & 76.11 & 76.11 & 18.33\dd{$-57.78$} \\
\quad Unsafe & 46.11 & 10.56\dd{$-35.56$} & 0.56\dd{$-45.56$} \\
\quad Task completed & 6.67 & 5.56\dd{$-1.11$} & 56.11\dd{$+49.44$} \\
\bottomrule
\end{tabular*}
\end{table}

\subsubsection{Runtime intervention on the alarm}
\label{sec:intervention}
On System~3, 42.88\% of the trajectories leave $\mathcal S$ within the horizon and at $K_{\max}\ln a=4$. Stopping at the alarm lowers the unsafe trajectories to 22.69\% (Table~\ref{tab:update-effects}). Without the online monitor and stop, no fault-free trajectory raises an alarm, 83 of the 180 faulty rollouts leave $\mathcal S$. With the stop, only 19 do.  Figure~\ref{fig:libero-frames} follows one rollout (wine-bottle task, $180^\circ$ rotation): without the stop the arm leaves $\mathcal S$ at $t=206$. With online monitor and intervention, DAB alarms at $t=154$ and the arm stays inside $\mathcal S$ until $K_{\max}=300$. We have also observed that not all unsafe trajectories will be alarmed in advance. This situation most often occurs on the drawer task under the 25\,N force and on the bowl task under the $137^\circ$ and $180^\circ$ rotations, where the robot arm leaves $\mathcal S$ within 35 to 39 steps, faster than the system can accumulate enough evidence (see Appendix~\ref{app:libero}, Table~\ref{tab:vla-faults}).

We should also point out honestly that one of the costs of intervention is the potential disruption of otherwise normal trajectories. Since in principle, we could trigger an alarm for every single trajectory and then intervene. However, we can manage this by adjusting $a$ the trade-off between intervention and normal operation. In the VLA task, with the alarm at $K_{\max}\ln a=-1$ and the stop alone, 5.56\% of the faulty trajectories complete the task, against 6.67\% without the stop, while the unsafe trajectories fall from 46.11\% to 10.56\%. At $a=1$ the stop leaves 22.22\% unsafe trajectories with the same 5.56\% completed. The stop mechanism sacrifices task completion for safety. To restore the ability to complete tasks, the policy needs to be updated. Therefore, Subsection~\ref{sec:retraining} updates the policy with the collected trajectories.

\subsubsection{Barrier and Policy Updates from Collected Data}
\label{sec:retraining}
The update step of Algorithm~\ref{alg1} retrains the barrier, the policy, or both from the collected episodes. The monitor depends on the policy and the barrier together. On System~2+input only the barrier is retrained, since the system has no controller to update. On System~3 the barrier and the policy are retrained together. After the update no trajectory raises an alarm on either system ($\hat p^{\max}=0.00\%$ and $\rho_\alpha=0$ on the same sampled initial states). System~2+input never leaves $\mathcal S$, before or after the update: its alarms come from the barrier condition alone and the retrained barrier removes them. On System~3 the unsafe trajectories fall from 42.88\% to 0.04\% (Table~\ref{tab:update-effects}). For the VLA, we adapt the deployed policy by correcting its action outputs. Over the faulty trajectories, the update raises the completed tasks from 6.67\% to 56.11\% and lowers the unsafe trajectories from 46.11\% to 0.56\% and the trajectories with an alarm from 76.11\% to 18.33\% (Table~\ref{tab:update-effects}). Only one completed trajectory raises an alarm.

Appendix~\ref{app:libero} repeats all three runs at $a=1$ and $K_{\max}\ln a=4$ (Table~\ref{tab:vla-a}). A larger $a$ alarms later, so the stop prevents fewer exits: with the stop, the unsafe trajectories rise from 10.56\% to 22.22\% at $a=1$ and to 36.11\% at $K_{\max}\ln a=4$. Of the three settings, $K_{\max}\ln a=-1$ leaves the fewer unsafe trajectories with the stop, which is significant on the robot VLA task.

\section{Conclusions and Future Work}
\label{sec:con}
We presented an anytime-valid framework for runtime monitoring of learned barrier certificates in stochastic systems. By constructing e-processes from observed barrier-value increments, the framework provides finite-sample false-alarm control under arbitrary stopping times. We developed two betting strategies, Direct and DAB, and introduced temporal boundary shaping to adjust alarm sensitivity over a prescribed deployment horizon. Experiments on benchmark systems and closed-loop robotic tasks demonstrated the framework's effectiveness in detecting statistical evidence against barrier conditions, supporting runtime intervention, and facilitating policy refinement.

Several directions remain for future work. First, extending the framework to safe reinforcement learning and partially observable systems would broaden its applicability. Another important direction is its extension to continuous-time stochastic systems. A key challenge in this setting is to detect boundary-hitting events along continuous trajectories. In particular, a trajectory may touch or cross the boundary of a safety-relevant set between two sampling instants and return before the next observation, causing conventional time-discretized monitoring to miss the event. Developing monitoring methods that can account for such inter-sample boundary-hitting events is therefore an important direction for future research. Second, the relationship between barrier-condition violations and detection power, detection delay, and pre-violation detection warrants further theoretical and empirical study. Moreover, the finite-horizon factor $a$ can be further studied as a time-varying parameter to adapt the monitoring strategy.

\noindent\textbf{Acknowledgments.} This work was partially supported by  National Research Foundation, Singapore, under the NRF RSS Scheme NRF-RSS2022-009, and Basic Research Program of Institute of Software, CAS (Grant No. ISCAS-JCMS-202302).

\FloatBarrier
\label{paper:body-end}
\clearpage
\bibliographystyle{\here ACM-Reference-Format}
\bibliography{\here references}

\clearpage
\appendix
\section*{Appendices}

\section{Proof and Benchmark details}
\label{app:benchmarks}
\subsection{Proof details}
\begin{proposition*}
Let $H_0$ be as in \eqref{eq:null-trajectories}. Suppose that $(M_t)_{t\ge0}$ is a nonnegative test supermartingale under $H_0$. Let $a\ge1$, $K_{\max}\in\mathbb{N}$, and $\alpha\in(0,1)$. Define the event $\mathcal R\coloneqq\{\tau_a\le K_{\max}\}$. Then
$\mathbb P_{H_0}(\mathcal R)\le\alpha$.
\end{proposition*}
\begin{proof}
Define the scaled process $V_t
:=
\frac{\alpha}{a^{K_{\max}}}M_t^a$. Since $(M_t)_{t\geq 0}$ is a nonnegative test supermartingale under
$H_0$,
\begingroup
\predisplaypenalty=0
\[
\begin{aligned}
\mathbb{E}_{H_0}[V_{t+1}\mid\mathcal{F}_t]
&=
\frac{\alpha}{a^{K_{\max}}}
a^{t+1}
\mathbb{E}_{H_0}[M_{t+1}\mid\mathcal{F}_t] \\
&\leq
\frac{\alpha}{a^{K_{\max}}}
a^{t+1} M_t = a V_t.
\end{aligned}
\]
\endgroup
Thus, $V_t$ satisfies the finite-horizon growth condition
\[
\mathbb{E}_{H_0}[V_{t+1}\mid\mathcal{F}_t]
\leq
\frac{V_t}{\bar{\alpha}}+\bar{\beta}
\]
from Theorem~1 in~\cite{xue2025finite}, with $\bar{\alpha}=\frac{1}{a}\in(0,1],\bar{\beta}=0$.
Moreover, $V_0 = \frac{\alpha}{a^{K_{\max}}},\bar{\gamma} =\bar{\beta}\bar{\alpha} -(\bar{\alpha}-1) = 1-\frac{1}{a} \in[0,1]$. Therefore, the case $\bar{\alpha}\in(0,1]$ and
$\bar{\gamma}\in[0,1]$ of Theorem~1 in~\cite{xue2025finite}
gives
\[
\begin{aligned}
\mathbb{P}_{H_0}\!\left(
\sup_{0\leq t\leq K_{\max}} V_t \geq 1
\right)\leq
V_0\bar{\alpha}^{-K_{\max}} =
\frac{\alpha}{a^{K_{\max}}}
a^{K_{\max}}=
\alpha.
\end{aligned}
\]
Finally,
\[
V_t\geq 1
\quad\Longleftrightarrow\quad
M_t^a\geq \frac{a^{K_{\max}}}{\alpha}
\quad\Longleftrightarrow\quad
M_t\geq
\frac{a^{K_{\max}-t}}{\alpha}.
\]
Hence,$\left\{
\sup_{0\leq t\leq K_{\max}}V_t\geq 1
\right\}
=
\{\tau\leq K_{\max}\}$, and therefore
\[
\mathbb{P}_{H_0}\!\left(\tau\leq K_{\max}\right)
\leq \alpha.
\]
This completes the proof.
\end{proof}

\begin{theorem*}[Stepwise falsification guarantee]
\label{thm:stepwise0}
Let $H_0$ be as in \eqref{eq:null-trajectories}. Suppose that $(M_t)_{t\geq0}$ is a nonnegative test supermartingale under $H_0$. Let $a>0$, $K_{\max}\in\mathbb N$ with $K_{\max}\geq1$, and $\alpha\in(0,1)$. Then the following hold:
\begin{enumerate}[label=(\roman*),leftmargin=*,nosep]
\item \emph{Weighted false-alarm budget.} The distribution
of the first alarm time satisfies
\[
\sum_{s=1}^{K_{\max}}
a^{K_{\max}-s}\,
\mathbb P_{H_0}(\tau_a=s)\leq\alpha.
\]

\item \emph{Cumulative false-alarm bound.} For every
$1\leq t\leq K_{\max}$,
\[
\mathbb P_{H_0}(\tau_a\leq t)\leq
\begin{cases}
\alpha\,a^{-(K_{\max}-t)}, & a\geq1,\\
\min\{1,\alpha\,a^{-(K_{\max}-1)}\}, & 0<a<1.
\end{cases}
\]
\end{enumerate}
In particular, over the whole horizon,
\[
\mathbb P_{H_0}(\tau_a\leq K_{\max})
\leq
\min\left\{1,\alpha\max\{1,a^{1-K_{\max}}\}\right\}.
\]
For $a\geq1$, the false-alarm level remains $\alpha$. Moreover, when $a>1$, the cumulative false-alarm bound is strictly below $\alpha$ at every step $t<K_{\max}$. In contrast, for $0<a<1$, the whole-horizon bound becomes $\min\{1,\alpha a^{1-K_{\max}}\}$, which is strictly below one exactly when
\[
(K_{\max}-1)\ln(1/a)<\ln(1/\alpha).
\]
\end{theorem*}

\begin{proof}
All probabilities and expectations below are under $H_0$. For (i), define the bounded stopping time
\[
T:=\min\{\tau_a,K_{\max}\}.
\]
By the optional stopping theorem for nonnegative supermartingales,
\[
\mathbb E_{H_0}[M_T]\leq M_0=1.
\]

By the definition of $\tau_a$, the events $\{\tau_a=s\}$, $s=1,\ldots,K_{\max}$, and $\{\tau_a=\infty\}$ form a partition of the sample space. On the event $\{\tau_a=s\}$, the stopped value satisfies $M_T=M_s\geq\frac{a^{K_{\max}-s}}{\alpha}$. On the event $\{\tau_a=\infty\}$, no alarm occurs within the monitoring horizon, so $T=K_{\max}$ and
$M_T=M_{K_{\max}}\geq0$, by the nonnegativity of the test supermartingale. Decomposing the expectation over these disjoint events gives
\[
\begin{aligned}
1
&\geq\mathbb E_{H_0}[M_T]\\
&=
\sum_{s=1}^{K_{\max}}
\mathbb E_{H_0}
\left[M_s\mathbf{1}_{\{\tau_a=s\}}\right]
+
\mathbb E_{H_0}
\left[M_{K_{\max}}
\mathbf{1}_{\{\tau_a=\infty\}}\right]\\
&\geq
\frac{1}{\alpha}
\sum_{s=1}^{K_{\max}}
a^{K_{\max}-s}
\mathbb P_{H_0}(\tau_a=s).
\end{aligned}
\]
Multiplying both sides by $\alpha$ establishes (i).

For (ii), since the possible first alarm times are mutually disjoint, $\mathbb P_{H_0}(\tau_a\leq t)=\sum_{s=1}^{t}\mathbb P_{H_0}(\tau_a=s)$. By (i), the corresponding partial sum satisfies $\sum_{s=1}^{t} a^{K_{\max}-s}\mathbb P_{H_0}(\tau_a=s) \leq\alpha$.

If $a\geq1$, then for every $1\leq s\leq t$, $a^{K_{\max}-s}\geq a^{K_{\max}-t}$. Consequently,
\[
\begin{aligned}
a^{K_{\max}-t}\mathbb P_{H_0}(\tau_a\leq t)\leq
\sum_{s=1}^{t}
a^{K_{\max}-s}\mathbb P_{H_0}(\tau_a=s) \leq\alpha.
\end{aligned}
\]
Dividing by $a^{K_{\max}-t}$ gives $\mathbb P_{H_0}(\tau_a\leq t)
\leq\alpha a^{-(K_{\max}-t)}$.

If $0<a<1$, then for every $1\leq s\leq t$, $a^{K_{\max}-s}\geq a^{K_{\max}-1}$.
Therefore,
\[
\begin{aligned}
a^{K_{\max}-1}\mathbb P_{H_0}(\tau_a\leq t)\leq
\sum_{s=1}^{t}
a^{K_{\max}-s}\mathbb P_{H_0}(\tau_a=s)\leq\alpha,
\end{aligned}
\]
which implies $\mathbb P_{H_0}(\tau_a\leq t) \leq\alpha a^{-(K_{\max}-1)}$. Combining this inequality with the trivial bound $\mathbb P_{H_0}(\tau_a\leq t)\leq1$ establishes the second case of (ii).
\end{proof}

\subsection{Benchmark details}
All the code and run records will be released with the artifact. A function $B$ is a barrier certificate if it satisfies the three conditions \eqref{eq:barrier-a}, \eqref{eq:barrier-b}, and \eqref{eq:barrier-supermartingale}. For a state $x\in\mathcal S$, let $x^+$ denote its next state. For $x\notin\mathcal S$, the stopped system stays at $x$, so \eqref{eq:barrier-supermartingale} holds with equality there, and it only has to be checked on $\mathcal S$. To show that a candidate is valid, all three conditions must hold everywhere. To show that it is invalid, we can just find a single \emph{counterexample} $x\in\mathcal S$ suffices: $\mathbb E[B(x^+)\mid x]>B(x)$.

\paragraph{System1: stochastic polynomial system (1D).}
\emph{Dynamics.}
The state $x\in\mathbb R$ evolves according to
\begin{equation*}
x_{k+1} =(0.5+1.5\xi_k)x_k(1-x_k^2),
\qquad
\xi_k\sim\mathcal U[-1,1].
\end{equation*}
The disturbances are independent across time.

\noindent\emph{Sets and sampling.}
\begin{align*}
\mathcal X_0
&=[-0.4,0.4],\;\;\mathcal S =(-1,1).
\end{align*}
Initial states are sampled uniformly from $\mathcal X_0$, independently of the disturbance sequence.

\paragraph{System~1: valid.}
We choose the candidate $B(x)=x^2$. It satisfies \eqref{eq:barrier-a}, and
\eqref{eq:barrier-b} holds since $|x|\geq 1$ implies $x^2\geq 1$. For \eqref{eq:barrier-supermartingale}, $x^+=(0.5+1.5\xi)\,x(1-x^2)$ with $\xi\sim\mathcal U[-1,1]$, and $\mathbb E[(0.5+1.5\xi)^2]=0.25+1.5\,\mathbb E[\xi]+2.25\,\mathbb E[\xi^2]=0.25+0+0.75=1$. Hence, for every
$x\in\mathcal S$,
\[
\begin{aligned}
\mathbb E[B(x^+)\mid x]-B(x)&=x^2(1-x^2)^2-x^2\\
&=-x^4(2-x^2)\le 0 .
\end{aligned}
\]

\paragraph{System2:  Van der Pol oscillator (2D)~\cite{wu2024framework}.}
\emph{Dynamics.}
The state $x=(x_1,x_2)\in\mathbb R^2$ evolves:
\begin{equation*}
\begin{aligned}
x_{1,k+1} &= x_{1,k}+0.01\,(-2x_{2,k}),\\
x_{2,k+1} &= x_{2,k}+0.01\big((0.8+\xi_k)\,x_{1,k}\\
&\qquad+10\,(x_{1,k}^2-0.21)\,x_{2,k}\big),
\end{aligned}
\end{equation*}
where $\xi_k\sim\mathcal U[-0.1,0.1]$. The disturbances are independent across time.

\noindent\emph{Sets and sampling.}
\begin{align*}
\mathcal X_0=\{x:\ x_1^2+x_2^2<0.01\},\;\;\mathcal S=\{x:\ x_1^2+x_2^2\le1\}.
\end{align*}
Initial states are sampled uniformly from $\mathcal X_0$, independently of the disturbance sequence.

\paragraph{System~2: valid.}
The candidate is a polynomial $B$ of degree $10$, computed together with SOS multipliers $s_0,s_1,s_2$ of degree $10$ by the SOS program using the method proposed in \cite{wu2024framework}:
\[
\begin{aligned}
\min_{B,\,s_0,\,s_1,\,s_2,\,\gamma}\ \ &\gamma\\
\text{s.t.}\ \ &B,\ s_0,\ s_1,\ s_2\ \text{are SOS},\\
&B-1-s_2\,(\|x\|^2-1)\ \text{is SOS},\\
&B-\mathbb E[B(x^+)\mid x]+s_1\,(\|x\|^2-1)\ \text{is SOS},\\
&\gamma-B+s_0\,(\|x\|^2-0.01)\ \text{is SOS}.
\end{aligned}
\]
Each constraint implies one condition. A sum of squares is nonnegative, so $B\geq 0$, which is \eqref{eq:barrier-a}. For $x\notin\mathcal S$, $\|x\|^2-1>0$ and $s_2\geq 0$, so $B(x)-1\geq s_2(x)(\|x\|^2-1)\geq 0$, which is
\eqref{eq:barrier-b}. For $x\in\mathcal S$, $\|x\|^2-1\le 0$ and $s_1\geq 0$, so $B(x)-\mathbb E[B(x^+)\mid x]\geq s_1(x)(1-\|x\|^2)\geq 0$, which is \eqref{eq:barrier-supermartingale}. We solve the program with YALMIP and MOSEK.  %

To simulate the changed environments in reality, we actively injected noise $b$ into state $x_2$ from step $k=100$ onward: 
\begin{equation}
    \begin{aligned}
        x_{2,k+1}=x_{2,k}&+0.01\big((0.8+\xi_k)\,x_{1,k}\\
        &+10\,(x_{1,k}^2-0.21)\,x_{2,k}+b_k\big)
    \end{aligned}
\end{equation}
where $b_k\sim\mathcal U[0,0.45]$ and the equation of $x_1$ is unchanged.

\paragraph{System~2+input: invalid.}
The candidate of system~2+input is the certificate of System~2.  Since both the system and the candidate barrier function are polynomial, we can directly compute the function value at a given point, as well as the $\mathbb E B(x^+)$. At $x=(-0.07,\,0.30)$, we obtain $\mathbb E[B(x^+)\mid x]=0.203196$ and $B(x)=0.198623$, so $x$ is a counterexample. Then the candidate in system~2 before is invalid in system~2+input.

\paragraph{System3: polynomial oscillator (2D)~\cite{abate2021fossil,vertovec2025scalable}.}
\emph{Dynamics.}
We applied Euler discretization to the continuous system from the original text. With $x=(x_1,x_2)$ and $\Delta t=0.1$, the system is
\begin{equation}
\begin{aligned}
x_{k+1}
&=x_k+\Delta t
\begin{bmatrix}
x_{2,k}\\
x_{1,k}^3/3-x_{1,k}-x_{2,k}+u_k
\end{bmatrix}
+\sigma w_k,\\
w_k
&\sim\mathcal U([-1,1]^2).
\end{aligned}
\end{equation}
The disturbances are independent across time and $\sigma = 0.01$. In the refinement stage, we train the policy and barrier function using neural networks with $|u|\leq1.5$. 

\noindent\emph{Sets and sampling.}
Let
\[
\begin{aligned}
\mathcal X_0
&=\mathcal B((1.5,0),0.5)
 \cup[-1.8,-1.4]\times[-0.1,0.1]\\
&\quad\cup[-1.4,-1.2]\times[-0.5,0.1],\\
\mathcal S&=[-3.5,2]\times[-2,1]\setminus \mathcal B((-1,-1),0.4)\\
&\quad\cup([0.4,0.6]\times[0.1,0.5])\\
&\quad\cup([0.4,0.8]\times[0.1,0.3]).
\end{aligned}
\]
where $\mathcal B((x_0,y_0),r)=\{(x,y)|(x-x_0)^2+(y-y_0)^2\leq r^2\}$.

\paragraph{System~3: invalid.}
The policy $u(x)=1.5\tanh(N_\pi(x))$ and the barrier $B(x)=\mathrm{softplus}(N_B(x))$ are neural networks trained from samples. Finding a valid certificate for such a stochastic system is difficult, and it is not needed here. The closed-loop dynamics are $x^+=m(x)+0.01\,w$ with
$w\sim\mathcal U([-1,1]^2)$, where
\[
m(x)=\bigl(x_1+0.1\,x_2,\;\;x_2+0.1\,(x_1^3/3-x_1-x_2+u(x))\bigr)
\]
is the next state without noise. Since $B$ is a neural network, it's hard to provide the closed form for $\mathbb E[B(x^+)\mid x]$. So we bound
it by partitioning the noise, as in \cite{cao2025comparative}, and by interval bound propagation through $N_B$~\cite{gowal2019effectivenessintervalboundpropagation}. We split $[-1,1]^2$ into $20\times 20$ cells $C_j$ of probability $1/400$ each. For $w\in C_j$, the successor lies in the box $Z_j=m(x)+0.01\,C_j$, and interval bound propagation through $N_B$ gives a lower bound $\underline B_j\le\min_{z\in Z_j}B(z)$. Hence
\[
\mathbb E[B(x^+)\mid x]\;\ge\;\frac{1}{400}\sum_{j=1}^{400}\underline B_j .
\]
The upper bound is obtained in the same way. At $x=(0.79,\,-0.32)$, we obtain
\[\mathbb E[B(x^+)\mid x]\in[0.132798,\,0.136322]\]
and $B(x)\le 0.126608$. The lower bound of the expectation exceeds $B(x)$, so $x$ is a counterexample and the candidate barrier function is invalid.

\paragraph{Computing environment.}
The experiments on Systems~1--3 were run on an Apple M4 Pro machine (12 CPU cores, 24\,GB of memory) with Python~3.12 (NumPy~2.4, SciPy~1.16, and PyTorch~2.5 for training). The SOS program for System~2 was solved with MATLAB R2025a, YALMIP and MOSEK~11.1. The VLA experiments were run on a Linux server with three NVIDIA A100 GPUs (40\,GB each): $\pi_{0.5}$ runs on a GPU through LeRobot, while the LIBERO simulator, the barrier and the DAB monitor run on the CPU.

\begin{figure*}[t]
    \centering
    \includegraphics[width=\textwidth]{\here 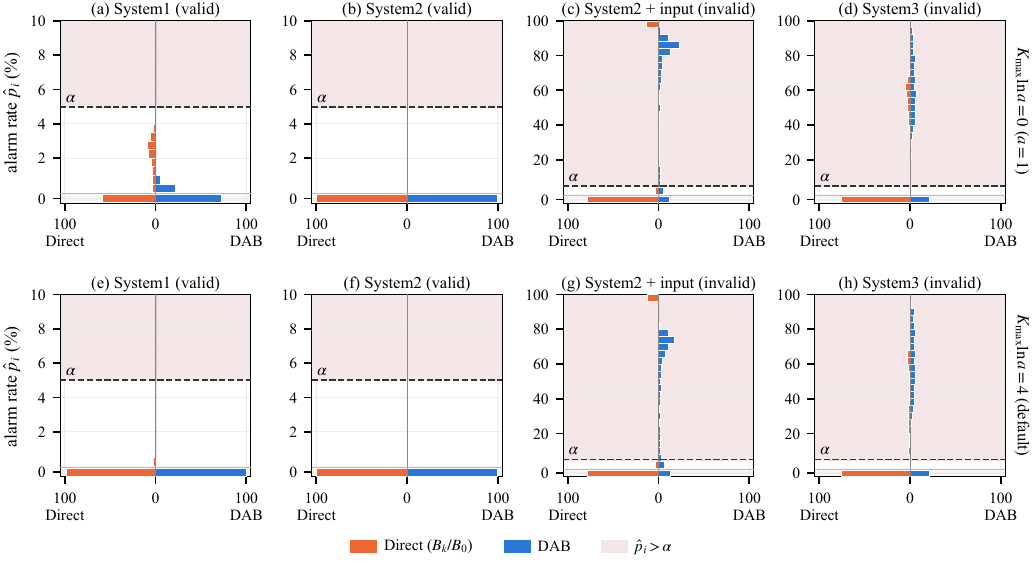}
    \caption{Distribution of the alarm rate $\hat p_i$ over the $N=10^3$ initial states, Direct (left half) against DAB (right half), with the standard rule $K_{\max}\ln a=0$ (top row) and the default $K_{\max}\ln a=4$ (bottom row); the row below each axis holds the share of initial states with $\hat p_i=0$.}
    \Description{Empirical alarm-rate distributions across sampled initial states, comparing Direct and DAB for the standard and temporally shaped rules.}
\label{fig:alarmrate}
\end{figure*}

\section{Further analysis}
\label{app:baselines-full}

\subsection{Additional Analysis of the Main Experiments}

\begin{figure*}[t]
    \centering
    \includegraphics[width=\textwidth]{\here 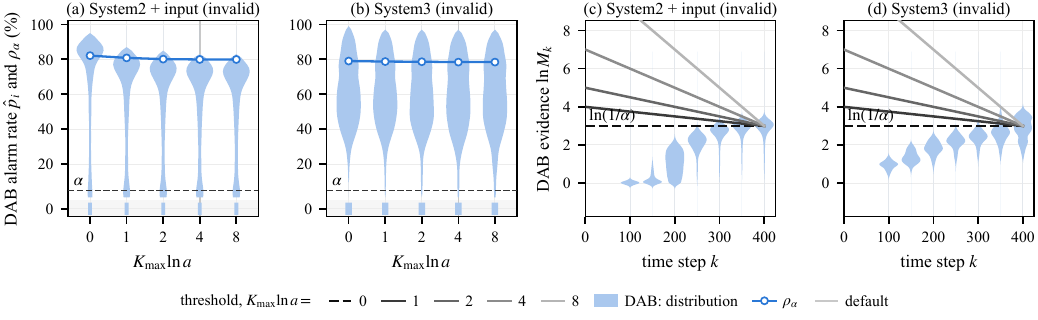}
    \caption{Effect of the patience parameter $a$ on the DAB alarm rates and evidence for the invalid certificates.}
    \Description{Alarm-rate summaries and accumulated evidence for different values of the temporal shaping parameter.}
\label{fig:factor_a}
\end{figure*}

\begin{figure*}[t]
    \centering
    \includegraphics[width=\textwidth]{\here 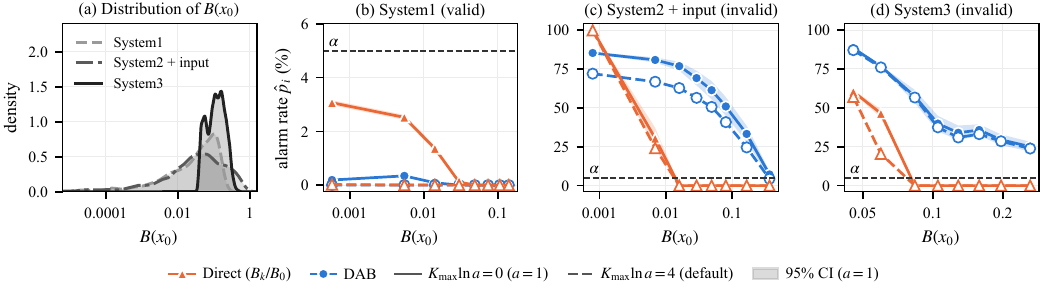}
    \caption{Initial barrier value $B(x_0)$ and alarm rate: solid lines and filled markers for the standard rule $K_{\max}\ln a=0$, dashed lines and hollow markers for the default $K_{\max}\ln a=4$.}
    \Description{Relationships between initial barrier values and empirical alarm rates for Direct and DAB.}
\label{fig:alarmrate2}
\end{figure*}

\noindent \textbf{Why does a change in $a$ have different effects on system~2 +input versus system~3?} Figure~\ref{fig:factor_a}(a, b) shows the distribution of DAB's alarm rates over the sweep $K_{\max}\ln a\in\{0,1,2,4,8\}$: $\rho_\alpha$ stays within $2.2$ points on both systems, while $\hat p^{\max}$ falls from $92.0\%$ to $81.0\%$ on System~2-input and only from $96.3\%$ to $95.0\%$ on System~3. Figure~\ref{fig:factor_a}(c, d) shows why. Every threshold equals $\ln(1/\alpha)$ at $k=K_{\max}$, so a run whose evidence is still above $\ln(1/\alpha)$ at the horizon raises an alarm for every $a$. A larger $a$ can only remove the alarm of a run that crosses $\ln(1/\alpha)$ before $K_{\max}$ and stays below the tilted thresholds. On System~2+input the evidence settles just above $\ln(1/\alpha)$ from about $k=300$ on, so many runs cross $1/\alpha$ late and only marginally, and the tilted thresholds remove part of them. On System~3 the evidence keeps rising until $K_{\max}$, so almost every run that crosses $1/\alpha$ also crosses the tilted thresholds.

\noindent \textbf{Why do the two betting strategies differ on the invalid systems?} Figure \ref{fig:alarmrate2} helps explain the different behaviors of the two betting strategies through the distribution of the initial barrier values $B(x_0)$. For direct betting, the evidence process satisfies $M_k=B(\hat{x}_k)/B(x_0)$, so the alarm condition for $a=1$ is equivalent to 
\[B(\hat{x}_k) \geq \frac{B(x_0)}{\alpha}.\]
Thus, a smaller initial barrier value yields a lower alarm threshold, making an alarm easier to trigger. System 2-input attack contains many initial states with small $B(x_0)$, and direct betting reaches a $100\%$ empirical alarm rate for initial states with $B(x_0)=0.001$. However, this behavior is concentrated among a subset of the sampled initial states, as reflected by $\rho_{\alpha}=17.90\%$ (Table~\ref{tab:validity}, $K_{\max}\ln a=4$; $18.10\%$ with $a=1$). The same logic applies to the case where $a > 1$, since the threshold $a^{K_{\max}-k}B(x_0)/\alpha$ is still proportional to $B(x_0)$.

In contrast, the initial barrier values in System 3 are generally larger, resulting in higher alarm thresholds for direct betting. Accordingly, Figure \ref{fig:alarmrate2}(b)--(d) shows that its empirical alarm rate decreases rapidly as $B(x_0)$ increases. DAB, however, adapts its betting fraction based on the observed barrier increments and exhibits nonzero empirical alarm rate over a broad range of initial barrier values. This helps explain why DAB achieves a higher fraction of empirically abnormal initial states and a higher worst-case empirical alarm rate than direct betting in System 3. 

\subsection{Sensitivity analysis}
\label{sec:exp-a}

Next, we study how the horizon $K_{\max}$, and the betting rule $\lambda_t$ affect the alarm rates. In these analyses, we vary only one parameter while keeping the others constant.

\begin{figure*}[t]
    \centering
    \includegraphics[width=\textwidth]{\here 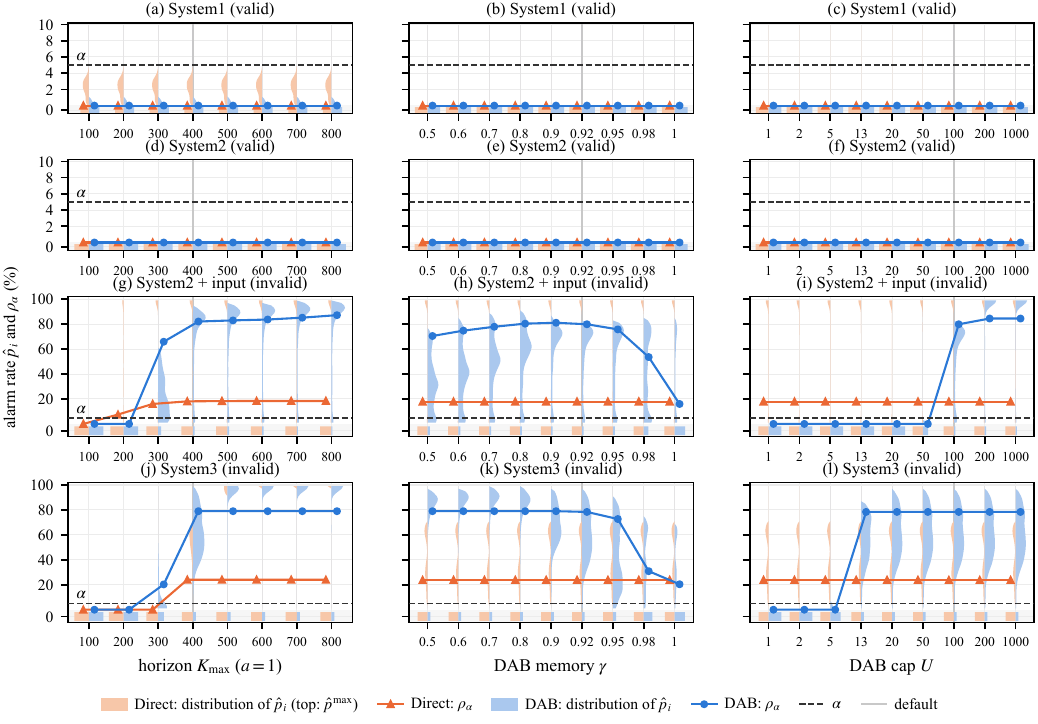}
    \caption{Sensitivity to the horizon $K_{\max}$, DAB memory $\gamma$ and DAB cap $U$.}
    \Description{Alarm-rate sensitivity to evaluation horizon, discount factor, and betting cap across valid and invalid certificate scenarios.}
\label{fig:sensitivity}
\end{figure*}

\noindent \textbf{Valid certificates.} No choice of these parameters can raise the false-alarm probability above $\alpha$. Under $H_0$, $\mathbb E[\delta_{t+1}\mid\mathcal F_t]\le 0$. The barrier can still increase
at single steps, but for every $\lambda_t$ chosen from past observations, $E_{t+1}\geq 0$ and $\mathbb E[E_{t+1}\mid\mathcal F_t]=1+\lambda_t\,\mathbb E[\delta_{t+1}\mid\mathcal F_t]\le 1$. Hence $M_k$ is a test
supermartingale for every $\gamma$ and $U$, and Theorem~\ref{thm:global} holds for every $K_{\max}$ and every $a\geq 1$. The first two rows of Figure~\ref{fig:sensitivity} confirm that the alarm rates of System~1 and System~2 stay below $\alpha$ in all settings. 

\noindent \textbf{Invalid certificates.} In scenarios where certificates are invalid, various parameters will influence the corresponding alarm rates, we will also provide specific recommendations for parameter selection in practical deployment.

\noindent  \paragraph{1. The horizon factor $K_{\max}$} 

We vary $K_{\max}\in\{100,200,\dots,800\}$ with $a=1$, i.e., with the constant threshold $1/\alpha$. We choose $a=1$ here not because it is the best value of $a$, but to isolate the effect of the horizon on the alarms: with $K_{\max}\ln a$ fixed, $a=e^{K_{\max}\ln a/K_{\max}}$ and the threshold at every step $k$ would change together with $K_{\max}$, so the effect of the horizon could not be separated from that of the threshold shaping, which is studied separately in Figure~\ref{fig:factor_a}. 

\textbf{Results.} Figure \ref{fig:sensitivity}(a,d) shows that the alarm rates for the valid systems remain unchanged as $K_{\max}$ varies.  With $a=1$ and a fixed $K_{\max}$, the share of runs that have raised an alarm by step $k$ is non-decreasing in $k$: as the monitored trajectory approaches $K_{\max}$, it can only increase or stay constant. The observed rates remain below or equal to $\alpha$, consistent with the anytime-valid guarantee. For System~2+input, the attack starts at $k=100$. Thus, when $K_{\max}=100$, the monitored trajectory contains no attacked step, and no alarms are raised (Figure~\ref{fig:sensitivity}(g)). For System~3, the barrier function is invalid from the start, but the evidence $\ln M_k$ needs about 300 steps to reach $\ln(1/\alpha)$. At $K_{\max}=100$, the evidence $\ln M_k$ remains well below the alarm threshold $\ln(1/\alpha)$ (Figure~\ref{fig:factor_a}(d)), so no alarms are raised (Figure~\ref{fig:sensitivity}(j)).  From $K_{\max}=300$ onward, DAB achieves higher $\rho_{\alpha}$ than Direct betting, indicating that DAB is more effective at detecting the loss of validity of the barrier certificates over these horizons. 

In practice, the choice of $K_{\max}$ involves a trade-off between the monitoring horizon and the desired level of confidence in the barrier certificate. A shorter horizon limits the time over which evidence against the certificate can accumulate, whereas a longer horizon provides more opportunities to detect a loss of validity. Since the validity of the barrier certificate is generally unknown a priori, $K_{\max}$ should be selected according to the practical monitoring requirements and the desired level of trust in the certificate. In particular, a shorter horizon may be appropriate when the certificate is considered highly trustworthy and limiting alarms is a priority, whereas a longer horizon may be preferred when more extensive monitoring is needed to detect potential invalidity.

\noindent \paragraph{2. Betting rule $\lambda_t$.} Direct uses $\lambda_t=1/B(\hat x_t)$ and has no tunable parameters. Therefore, its alarm rates remain unchanged across the parameter setting in the second and third column of the Figure~\ref{fig:sensitivity}. In contrast, DAB uses 
\[\lambda_t=\min(\lambda_t^{\mathrm K},U,1/B(\hat x_t)),\] where $\lambda_t^{\mathrm K}$ depends on the discounted sums $S_t$ and $Q_t$ of past increments, with the discount factor $\gamma$. When $\gamma=1$, earlier decreases in the barrier can offset subsequent increases in $S_t$. As shown in Figure~\ref{fig:sensitivity}(h,k), both invalid systems consequently exhibit substantially fewer alarms. Lowering $\gamma$ from 1 raises detection on both systems. Below 0.9, System~3 is flat while System~2+input declines, indicating that the effectiveness of the discount factor depends on the system. 

The cap $U$ mainly limits the betting factor when the Kelly-based factor is not the active constraint. %
Figure~\ref{fig:sensitivity}(i, l) shows that DAB produces alarm rates above $\alpha$ for System~2+input only for $U\geq 100$, whereas the alarm rates for System~3 change little once $U\geq 13$. These results suggest that the value of $U$ needed to achieve effective detection depends on the system. In practice, $U$ can be selected through a sensitivity analysis, choosing a value beyond which increasing the cap yields little additional improvement in detection performance. Increasing $U$ does not compromise the anytime-valid guarantee, because the constraint $\lambda_t\le 1/B(\hat x_t)$ remains in force. Consistent with this guarantee, the alarm rates for the valid systems remain below $\alpha$, even when $U=1000$.

We also try other betting strategies. We compare our adaptive betting rule with fixed betting rules computed from the same barrier increments. For any betting sequence with $0\leq\lambda_t\leq 1/B_t$, define $M_t(\lambda)=\prod_{s=1}^{t}(1+\lambda_{s-1}\delta_s)$. Monitors use the threshold $a^{K_{\max}-t}/\alpha$ of the main experiments ($K_{\max}\ln a=4$) for their own betting sequences, and they use the average of the $M_s(\lambda)$ values for multiple fixed parameter values when they have the same weights. For Direct, $M_t=B_t/B_0$. The fraction mixture uses $\lambda_t^{(c)}=c/B_t,\ c\in\mathcal C=\{0.2,0.5,0.75,1\}$. The absolute mixture uses the fixed absolute bet $\lambda_t^{(\beta)}$ for $\beta\in\mathcal A$ with $\mathcal A_1=\{1,10,20,100\}$, and $\mathcal A_2=\{10,30,60,100\}$. In each episode, each component of the absolute mixture keeps $c$ and $\beta$ fixed.  In DAB, $S_t$ and $Q_t$ are $\gamma$-discounted sums of past $\delta_s$ and $\delta_s^2$ (Eq.~\eqref{eq:safe-bet}). 

\begin{ResultBox}
\noindent\textbf{What do these methods guarantee?}
For betting components under $H_0$, their equal-weight mixtures are also nonnegative supermartingales. So Theorem~\ref{thm:global} can be applied to all betting rules in Table~\ref{tab:monitors}.
\end{ResultBox}

\begin{table}[t]
\caption{Compared betting rules, all driven by the increments $\delta_t=B_t-B_{t-1}$. Guar.: $P_{H_0}(\tau\le K_{\max}\mid x_0)\le\alpha$ for every $x_0$ (Theorem~\ref{thm:global}).}
\label{tab:monitors}
\centering
\footnotesize
\setlength{\tabcolsep}{3pt}
\renewcommand{\arraystretch}{1.25}
\begin{tabular}{@{}l l l c@{}}
\toprule
Betting rules & Bet $\lambda_t$ & Alarm & Guar. \\
\midrule
Direct & $1/B_t$ & $M_t\ge \frac{a^{K_{\max}-t}}{\alpha}$ & \checkmark \\
Fraction mix. & $c/B_t,\ c\in\mathcal C$ & $\bar M_t\ge \frac{a^{K_{\max}-t}}{\alpha}$ & \checkmark \\
Absolute mix. & $\min\{\beta,1/B_t\},\ \beta\in\mathcal A_j$ & $\bar M_t\ge \frac{a^{K_{\max}-t}}{\alpha}$ & \checkmark \\
\rowcolor{gray!12}
\textbf{DAB} & Eq.~\ref{eq:safe-bet} & $M_t\ge \frac{a^{K_{\max}-t}}{\alpha}$ & \checkmark \\
\bottomrule
\end{tabular}

\end{table}

\begin{figure*}[t]
   \centering
   \includegraphics[width=1\linewidth]{\here 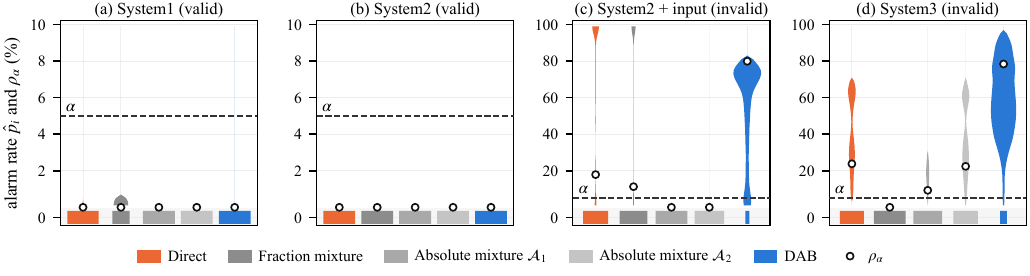}
   \caption{Distribution of the alarm rates $\hat p_i$ over the initial states for all monitors ($K_{\max}\ln a =4$).}
   \Description{Empirical alarm-rate distributions comparing Direct, fixed-bet mixtures, and DAB across the benchmark scenarios.}
\label{fig:baselines}
\end{figure*}

\noindent \textbf{Results.} Table~\ref{tab:baselines} and Figure~\ref{fig:baselines} show that all monitors keep the alarm rates of the valid systems below $\alpha$, with $\hat p^{\max}\le 0.50\%$ on System~1 and no alarm on System~2. 

On the invalid systems, DAB raises alarm rates above $\alpha$ for $79.90\%$ of the initial states of System~2-input attack and $78.40\%$ of those of System~3. For Direct, these shares are only $17.90\%$ and $23.80\%$, and for the mixtures they are at most $22.40\%$. This is because a fixed bet is also placed while the barrier decreases, which multiplies the evidence by a factor below one. For example, the barrier of System~2-input attack decreases for many steps before the attack, and the fixed bets lose evidence during this period. DAB does not bet when the discounted sum $S_t$ of recent increments is negative, and thus keeps its evidence for the steps where the barrier rises. Moreover, the mixture methods may be less effective because they average the evidence processes of their components. Consequently, strong evidence accumulated by a single component may be diluted by the others, preventing the mixture from reaching the alarm threshold. 

\begin{table*}[t]
\caption{Comparison with other monitors on the runs of
Table~\ref{tab:validity}.
$N=10^3$ initial states, $R=10^3$ trajectories per initial state,
$K_{\max}=400$, $K_{\max}\ln a=4$, $\alpha=0.05$.
All values are percentages.}
\label{tab:baselines}
\centering
\footnotesize
\renewcommand{\arraystretch}{1.1}
\begin{tabular*}{\linewidth}{@{\extracolsep{\fill}} l rr rr rr rr rr @{}}
\toprule
& \multicolumn{2}{c}{Direct}
& \multicolumn{2}{c}{Fraction mix.}
& \multicolumn{2}{c}{Absolute mix. $\mathcal A_1$}
& \multicolumn{2}{c}{Absolute mix. $\mathcal A_2$}
& \multicolumn{2}{c}{DAB} \\
\cmidrule(lr){2-3} \cmidrule(lr){4-5}
\cmidrule(lr){6-7} \cmidrule(lr){8-9} \cmidrule(l){10-11}
System
& $\rho_\alpha$ & $\hat{p}^{\max}$
& $\rho_\alpha$ & $\hat{p}^{\max}$
& $\rho_\alpha$ & $\hat{p}^{\max}$
& $\rho_\alpha$ & $\hat{p}^{\max}$
& $\rho_\alpha$ & $\hat{p}^{\max}$ \\
\midrule
\multicolumn{11}{l}{\textit{Valid certificates}} \\
System~1 & 0.00 & 0.20 & 0.00 & 0.50 & 0.00 & 0.00 & 0.00 & 0.00 & 0.00 & 0.10 \\
System~2 & 0.00 & 0.00 & 0.00 & 0.00 & 0.00 & 0.00 & 0.00 & 0.00 & 0.00 & 0.00 \\
\midrule
\multicolumn{11}{l}{\textit{Invalid certificates}} \\
System~2-input attack & 17.90 & 100.00 & 11.30 & 100.00 & 0.00 & 0.00 & 0.00 & 1.20 & 79.90 & 81.50 \\
System~3 & 23.80 & 69.40 & 0.00 & 0.00 & 9.30 & 29.50 & 22.40 & 68.00 & 78.40 & 95.10 \\
\bottomrule
\end{tabular*}

\end{table*}

\section{Training Details}

We train the neural barrier $B_\phi$ (with a softplus output layer so that $B_\phi \ge 0$ structurally) by minimizing soft penalties on the barrier conditions. Given a batch of constraint states $\{x_i\}_{i=1}^{N}$ sampled from the safe region, we simulate $M$ independent successor states $x_{i,j}' \sim P(\cdot \mid x_i, \pi_\theta(x_i))$ for each $x_i$ and estimate the conditional expectation by $\widehat{\mathbb{E}}[B_\phi(x') \mid x_i] = \frac{1}{M}\sum_{j=1}^{M} B_\phi(x_{i,j}')$.
The supermartingale decrease condition is enforced by
\begin{equation}
\mathcal{L}_{\mathrm{dec}} = \frac{1}{N} \sum_{i=1}^{N} \mathrm{ReLU}\!\Big( \widehat{\mathbb{E}}[B_\phi(x') \mid x_i] - B_\phi(x_i) \Big),
\end{equation}
while the initial and unsafe conditions are enforced on sampled sets $\{x_0^{(i)}\}_{i=1}^{N_0} \subset \mathcal{X}_0$ and $\{x_u^{(i)}\}_{i=1}^{N_u} \subset \mathcal{X}_u$ by
\begin{equation}
\begin{aligned}
\mathcal{L}_{\mathrm{init}} &= \frac{1}{N_0} \sum_{i=1}^{N_0} \mathrm{ReLU}\big( B_\phi(x_0^{(i)}) - b \big),\\
\mathcal{L}_{\mathrm{unsafe}} &= \frac{1}{N_u} \sum_{i=1}^{N_u} \mathrm{ReLU}\big( 1 - B_\phi(x_u^{(i)}) \big),
\end{aligned}
\end{equation}
where $b>0$ is an upper bound on the initial barrier value. The total loss is
\begin{equation}
\mathcal{L}(\phi) = w_1\,\mathcal{L}_{\mathrm{dec}} + w_2\,\mathcal{L}_{\mathrm{unsafe}} + w_3\,\mathcal{L}_{\mathrm{init}},
\label{eq:barrier-loss}
\end{equation}
where $w_1,w_2,w_3>0$ are weighting parameters. All terms are optimized with Adam. In this paper, we mainly focus on online training, where collecting multiple independent successors from the same state may be impractical. We therefore set $M=1$. Although the resulting single-sample Monte Carlo estimate is unbiased for $\mathbb{E}[B_\phi(x')\mid x_i]$, it has higher variance. Moreover, applying the nonlinear ReLU penalty to a single-sample estimate does
not provide an unbiased estimate of the ideal expected decrease loss.
Consequently, the learned barrier is not necessarily a valid certificate. Nevertheless, the runtime monitor can detect potential violations and intervene against unsafe behavior. 
The neural network output $B_\phi$ uses a Softplus activation, which
ensures that $B_\phi(x)\geq 0$. We then define the deployed barrier
function by the piecewise extension
\begin{equation}
\bar{B}_\phi(x)
=
\begin{cases}
B_\phi(x), & x\in\mathcal S,\\
1, & x\notin\mathcal S.
\end{cases}
\label{eq:piecewise-barrier}
\end{equation}
Therefore, the nonnegativity condition and the unsafe-set condition
$\bar{B}_\phi(x)\geq 1$ for all $x\notin\mathcal S$ are satisfied
structurally. During deployment, the runtime monitor evaluates
$\bar{B}_\phi$ and computes the barrier increments as
\begin{equation}
\delta_{t+1}
=
\bar{B}_\phi(\hat{x}_{t+1})
-
\bar{B}_\phi(\hat{x}_t).
\end{equation}
The sampled unsafe-set loss is retained as an auxiliary training term
to encourage the neural-network output to approach the prescribed
unsafe value near the boundary of the safe set.

A later alarm also leaves longer trajectories for the update: the completed tasks reach 62.22\% at $a=1$ and 59.44\% at $K_{\max}\ln a=4$, against 56.11\% at $K_{\max}\ln a=-1$. The choice of $a$ thus weighs the exits prevented by the stop against the data for the update.

\paragraph{Training settings.} All networks are trained with Adam and have two tanh hidden layers (32 units for the policy of System~3, 64 for every barrier). The policy of System~3 is $u=1.5\tanh(N_\pi(x))$, and every neural barrier has a softplus output. The initial barrier of System~2 is computed by SOS programming and is not trained. After the System~+2 input, we use the NN to train the barrier function. We set minimizes \eqref{eq:barrier-loss} with $M=1$.

\section{VLA details}
\label{app:libero}

We set $K_{\max}=300$ control steps (15\,s at 20\,Hz). One barrier per task is trained on 20 rollouts and frozen, and the monitor uses $\alpha=0.1$, $\gamma=0.92$, $U=100$. Each task uses ten test initial states of LIBERO-Goal. Each is run with a fixed random seed, the same in every run and for every $a$, so the runs of Tables~\ref{tab:update-effects} and~\ref{tab:vla-a} differ only in the stop, the update and~$a$. The trajectory data used for policy updates is drawn from three additional trajectories, in order to prevent any overlap between the test set and the training set.

\paragraph{Barrier training.} For each task we train one barrier $B_\phi$ with the loss~\eqref{eq:barrier-loss} and $M=1$, as for the benchmarks. The data are the transitions of 20 fault-free trajectories of the frozen policy, from initial states disjoint from the test and update ones, together with 120 constructed unsafe states. The input is the 134 physical coordinates of the simulator state (joint positions and velocities, end-effector pose, object poses), and $B_\phi$ has two hidden layers of 64 tanh units and a softplus output. We use $b=0.05$, $(w_1,w_2,w_3)=(10,1,1)$ and Adam (learning rate $10^{-3}$, batch 256, 2000 steps). The barrier is then frozen and used unchanged in every run.

We mainly consider the following scenario: after the robot receives images of the environment and the corresponding task specification, it generates commands and the associated control input $u_t$. However, the mapping from the control command $u_t$ to its actual execution is often affected by structural aging, noise and control execution errors, which may lead to dangerous trajectories. To address this issue, our online shield accumulates evidence based on the real operational state of the manipulator. In fact, the required data varies for different policy update methods and there are multiple alternatives for subsequent policy update. One is end-to-end fine-tuning of the entire policy using all collected data at each step, and the other is fine-tuning only the output control $u_t$. Considering the practical training cost, we adopt the second approach.

\paragraph{Policy update details.} Let $u_t$ be the planar command sent to the robot and $d_t$ the planar displacement of the end effector in step $t$. The true relation $d_t=\mathcal F(x_t,u_t)$ is unknown. On the 20 fault-free rollouts that train the barrier, we use the linear model $d_t\approx G_0u_t+\Lambda_0d_{t-1}+c_0$ to fit and  $R^2=0.99$. The update follows Algorithm~\ref{alg1} with $N_{\mathrm{batch}}=3$: each collected trajectory ends at its alarm or its exit, and $G$ and $c$ are fitted by least squares on all collected steps,
\begin{equation}
d_t-\Lambda_0 d_{t-1}=G\,u_t+c,
\label{eq:repair-fit}
\end{equation}
and in later episodes the policy sends
\begin{equation}
u_t^{\textbf{update}}=G^{-1}\big(G_0u_t^{\pi}-(c-c_0)\big).
\label{eq:repair-apply}
\end{equation}
Here, $u_t^\pi$ denotes the original planar policy output. The same fit is applied to each fault without information about the fault.

\begin{table}[h]
\caption{VLA trajectory counts per fault ($K_{\max}\ln a=-1$), the runs of Table~\ref{tab:update-effects}. Al.: trajectories with an alarm; Uns.: trajectories that leave $\mathcal S$; Succ.: trajectories that complete the task.}
\label{tab:vla-faults}
\centering
\footnotesize
\setlength{\tabcolsep}{2.5pt}
\begin{tabular}{@{}lrcccccccc@{}}
\toprule
 & & \multicolumn{2}{c}{No stop} & \multicolumn{3}{c}{Stop} & \multicolumn{3}{c}{Updated}\\
\cmidrule(lr){3-4}\cmidrule(lr){5-7}\cmidrule(l){8-10}
Fault & $n$ & Uns. & Succ. & Al. & Uns. & Succ. & Al. & Uns. & Succ.\\
\midrule
Fault-free & 30 & 0 & 26 & 0 & 0 & 26 & -- & -- & -- \\
Rotation $90^\circ$ & 30 & 12 & 0 & 24 & 2 & 0 & 1 & 0 & 23 \\
\quad $137^\circ$ & 30 & 15 & 0 & 25 & 5 & 0 & 6 & 0 & 9 \\
\quad $180^\circ$ & 30 & 24 & 0 & 25 & 5 & 0 & 14 & 0 & 11 \\
\quad $90^\circ$ + dead zone & 30 & 12 & 0 & 24 & 0 & 0 & 1 & 1 & 20 \\
Force 10\,N & 30 & 0 & 12 & 15 & 0 & 10 & 5 & 0 & 20 \\
\quad 25\,N & 30 & 20 & 0 & 24 & 7 & 0 & 6 & 0 & 18 \\
\midrule
Total (faults) & 180 & 83 & 12 & 137 & 19 & 10 & 33 & 1 & 101 \\
\bottomrule
\end{tabular}
\end{table}

\paragraph{The patience parameter on the VLA} Table~\ref{tab:vla-a} repeats the three runs(No stop, stop and updated) of Table~\ref{tab:update-effects} at $a=1$ and $K_{\max}\ln a=4$, with the update trajectories collected at the same $a$. Gray entries give the change from the main text ($K_{\max}\ln a=-1$). No setting raises an alarm on the fault-free trajectories. The stop results change monotonically with $a$: a larger $a$ raises the boundary at every step, so it cannot add an alarm or bring one forward (Subsection~\ref{exp:patience}). The trajectories with an alarm fall from 76.11\% to 35.00\% and the unsafe trajectories under the stop rise from 10.56\% to 36.11\%. At $K_{\max}\ln a=4$ the faults that move the arm out quickly leave $\mathcal S$ before the evidence reaches the higher boundary.

The test initial states and seeds are the same for every $a$. In the updated results, the alarm rate changes monotonically with $a$ (18.33\%, 11.67\% and 6.67\%). But there are differences in the unsafe and task completed for the monotonically. This is because $a$ enters the update only through the point where each update trajectory is cut, which changes the fitted $G$ and $c$. Slightly different adjustments can alter the entire closed-loop trajectory. A longer fit is therefore not uniformly better: on the bowl task under the $90^\circ$ rotation the fit grows from 197 to 294 and 733 transitions while the successes go from 10 to 10 and 8. The one systematic effect is on the bowl task under the $137^\circ$ and $180^\circ$ rotations, where $K_{\max}\ln a=-1$ alarms after 23 to 38 steps and leaves 94 and 85 transitions for the fit; 1 of the 20 updated trajectories succeeds, against 16 at $a=1$ with 141 and 115 transitions. Raising $a$ is thus the way to give the update more data when the stop matters less than the repaired task. From this, it becomes clear that determining how to effectively utilize collected data for training barriers or policies—specifically, which data to use and how to configure the parameters for training updates—remains a valuable area for research. Our online monitor offers an alternative perspective: it enables real-time intervention while simultaneously gathering the data required for model updates.

\begin{table}[h               ]
\caption{The three runs of Table~\ref{tab:update-effects} on the VLA at three values of $a$ (DAB, $\alpha=0.1$), in \% of the 180 faulty trajectories. Gray: change from $K_{\max}\ln a=-1$. Lead: median number of steps from the alarm to the exit, over the exits with an earlier alarm.}
\label{tab:vla-a}
\centering
\footnotesize
\setlength{\tabcolsep}{4pt}
\renewcommand{\arraystretch}{1.1}
\begin{tabular*}{\columnwidth}{@{\extracolsep{\fill}}lccc}
\toprule
$K_{\max}\ln a$ & $-1$ (main text) & $0$ ($a=1$) & $4$ \\
\midrule
\multicolumn{4}{l}{\emph{No stop}} \\
\quad Alarm rate & 76.11 & 72.78\dd{$-3.33$} & 35.00\dd{$-41.11$} \\
\quad Unsafe & 46.11 & 46.11 & 46.11 \\
\quad Task completed & 6.67 & 6.67 & 6.67 \\
\addlinespace
\multicolumn{4}{l}{\emph{Stop}} \\
\quad Unsafe & 10.56 & 22.22\dd{$+11.67$} & 36.11\dd{$+25.56$} \\
\quad Task completed & 5.56 & 5.56 & 6.67\dd{$+1.11$} \\
\addlinespace
\multicolumn{4}{l}{\emph{Updated}} \\
\quad Alarm rate & 18.33 & 11.67\dd{$-6.67$} & 6.67\dd{$-11.67$} \\
\quad Unsafe & 0.56 & 2.22\dd{$+1.67$} & 1.11\dd{$+0.56$} \\
\quad Task completed & 56.11 & 62.22\dd{$+6.11$} & 59.44\dd{$+3.33$} \\
\midrule
Lead (steps) & 52 & 64\dd{$+12$} & 50.5\dd{$-1.5$} \\
\bottomrule
\end{tabular*}
\end{table}

\end{document}